\documentclass[a4,semcolor,psfig,english,12pt,epsf,portrait]{article}
\usepackage{amsmath,amsfonts,latexsym,amscd,amssymb,theorem}

\usepackage[T1]{fontenc}
\usepackage{epsfig}
\usepackage{amsmath}
\usepackage{psfrag}

\def\div{\hbox{div}}

\def\R{\hbox{\bf R}}
\def\Z{\hbox{\bf Z}}

\def\div{\hbox{div}}

\def\T{\mathbb{ T}}

\def\a{\alpha}

\def\D{{\cal D}}

\def\e{\varepsilon}

\def\<{\langle}
\def\>{\rangle}
\newcommand{\ba}{\begin{eqnarray}}
\newcommand{\ea}{\end{eqnarray}}

\newtheorem{theo}{\bf Theorem}[section]
\newtheorem{lem}[theo]{\bf \textit{Lemma}}
\newtheorem{pro}[theo]{\bf Proposition}
\newtheorem{cor}[theo]{\bf Corollary}
\newtheorem{defi}[theo]{\bf Definition}
\newtheorem{rem}[theo]{\bf Remark}

\renewcommand{\R}{{\mathbb R}}
\renewcommand{\Z}{{\mathbb Z}}

\begin{document}

\title{\bf Global existence for a system of non-linear and non-local 
transport equations describing the dynamics of dislocation densities}

\author{
\normalsize\textsc{M. Cannone$^2$, A. El
  Hajj$^1$ $^2$, R. Monneau$^1$,
F. Ribaud$^2$
}}
\vspace{20pt}
\maketitle
\footnotetext[1]{\'Ecole Nationale des Ponts et
 Chauss\'ees, CERMICS,
6 et 8 avenue Blaise Pascal, Cit\'e Descartes
 Champs-sur-Marne, 77455 Marne-la-Vall\'ee Cedex 2, France}
\footnotetext[2]{ Université de Marne-la-Vallée
5, boulevard Descartes
Cité Descartes - Champs-sur-Marne 
77454 Marne-la-Vallée cedex 2}


 \centerline{\small{\bf{Abstract}}}
 \noindent{\small{In this paper, we study the global in time existence
     problem for the  Groma-Balogh model
     describing the dynamics of dislocation densities. This model is a
     two-dimensional model where the
     dislocation densities satisfy a system of transport
     equations such that the velocity vector field is the shear stress in
     the material, solving the equations of elasticity. This shear stress
     can be expressed  as some Riesz transform of the dislocation
     densities. The main tool in the proof of this result is the
     existence of an entropy for this system.}}

\hfill\break
 \noindent{\small{\bf{AMS Classification: }}} {\small{54C70, 35L45, 35Q72, 74H20, 74H25.}}\hfill\break
  \noindent{\small{\bf{Key words: }}} {\small{Cauchy's problem, system
      of non-linear transport equations, system
      of non-local transport equations, system of hyperbolic equations,
      entropy, Riesz transform, Zygmund space, dynamics of dislocation densities.}}\hfill\break


\section{Introduction}
\subsection{Physical motivation and presentation of the model}

Real crystals show certain  defects in the organization
of their crystalline structure, called dislocations. 
These defects were introduced in the Thirties by Taylor,
Orowan and Polanyi as the principal explanation of plastic
deformation at the microscopic scale of materials.\\

\noindent In a particular
case where these defects are parallel lines in the three-dimensional
space, their cross-section can be viewed as points in a plane. Under the
effect of an exterior stress, dislocations can be moved. In the special
case of what  is called ``edge dislocations'', these dislocations move in
the direction of their ``Burgers vector'' which has a fixed
direction. (cf  J. Hith and J. Lothe \cite{Hirth} for more physical
description).\\

\begin{figure}[h]\label{EC:fig:1}
\psfrag{X1}{ $\hspace{-1em}$$x_2$}
\psfrag{X2}{$x_1$}
\psfrag{m}{$\top$}
\psfrag{P}{$\bot$}
\psfrag{a}{{\scriptsize $-\vec{b}$}}
\psfrag{b}{{\scriptsize $+\vec{b}$}}
\psfrag{DISLO+}{\small  dislocation of $+$ type}
\psfrag{DISLO-}{\small $\hspace{-8em}$ dislocation of $-$ type}
\centering\epsfig{file=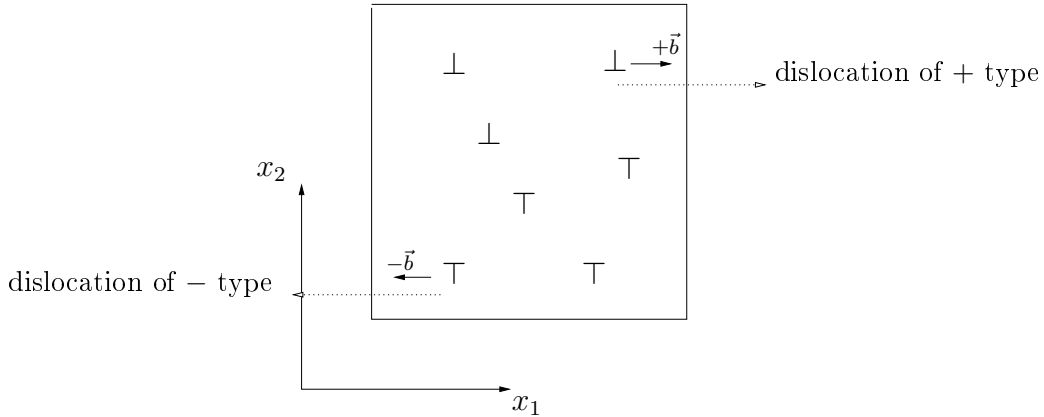, width=80mm}
\caption{Groma-Balogh 2D model.}
\end{figure}

\noindent In this work, we are interested in the mathematical study of a
model introduced by I. Groma, P. Balogh in \cite{Groma2} and
\cite{Groma}. In this model we consider two types of dislocations in
the plane $(x_1,x_2)$. Typically for a given velocity field, those dislocations of
type $(+)$ propagate in the direction  $+\vec{b}$ where $\vec{b}=(1,0)$
is the Burgers vector, while those of type $(-)$ propagate in the
direction $-\vec{b}$ (see Figure \ref{EC:fig:1}).\\

\noindent Here the velocity vector field is the shear stress in
 the material, solving the equations of elasticity. It turns out that this
 shear stress can be  expressed  as some Riesz transform of the solution
 (see Section \ref{EC:subsce:model}). More precisely our non-linear and
 non-local system of transport equations is the following:

\begin{center}\begin{equation}\tag{P}\label{EC:eq:i:1}\left\{\begin{array} {lll}
\displaystyle{\frac{\partial{\rho}^{+}}{\partial t}(x,t)} =-
&\left(R_1^2R_2^2\left(\rho^+(\cdot,t)-\rho^-(\cdot,t)\right)(x)\right)
\displaystyle{\frac{\partial{\rho}^{+}}{\partial
    x_1}(x,t)}&\mbox{in $\mathbb \D'(\R^2\times(0,T))$,}\\
 \\
\displaystyle{\frac{\partial{\rho}^{-}}{\partial
    t}(x,t)}  =
&\left(R_1^2R_2^2\left(\rho^+(\cdot,t)-\rho^-(\cdot,t)\right)(x)\right)
\displaystyle{\frac{\partial{\rho}^{-}}{\partial

    x_1}(x,t)}&\mbox{in  $\mathbb\D'(\R^2\times(0,T))$.}
\end{array} \right.\end{equation}\end{center}

\noindent The unknowns of the system (\ref{EC:eq:i:1}) are the scalar functions $\rho^{+}$ and
$\rho^{-}$ at the time $t$ and the position $x=(x_1,x_2)$, that we denote
for simplification by $\rho^{\pm}$. These terms correspond to the plastic
deformations in a crystal. Their derivative in the  $x_1$ direction
(i.e. the direction of  Burgers vector $\vec{b}$),
$\displaystyle{\frac{\partial{\rho}^{\pm}}{\partial x_1}}$ represents
the dislocation densities of $\pm$ type. In our work, we will only
consider solutions ${\rho}^{\pm}$ such that
$\displaystyle{\frac{\partial{\rho}^{\pm}}{\partial t}}$,
$\nabla{\rho}^{\pm}$ and ${\rho}^{+}-{\rho}^{-}$ are $\Z^2$-periodic
functions. The operators $R_1$ (resp. $R_2$)  are the
Riesz transformations associated to $x_1$ (resp. $x_2$). More precisely, these
Riesz transforms are defined as follows:

\begin{defi}\label{EC:defi:riesz}{\bf(Riesz transform in the periodic case)}\\
Let the torus $\T^2=\R^2/ \Z^2$. We define for $i\in \{1,2\}$  the Riesz
transforms $R_i$ over $\T^2$ as follows. If $f\in L^2(\T^2)$,  the
Fourier series coefficients of $R_if$ are given by:

\noindent i) $c_{(0,0)}\left(R_if\right)=0,$

\noindent ii) $c_k\left(R_if\right)=\displaystyle{\frac{k_i}{|k|}}c_k(f) \;\;\;
\mbox{for $k=(k_1,k_2)\in\mathbb Z^2\setminus\{(0,0)\},$}$

\noindent where  we recall that $\displaystyle{c_k(f)=\int_{\T^2}f(x)e^{-2\pi
    i k\cdot x}d^2x}$.
\end{defi}

\noindent In fact, this 2D model has been generalized later
 in 2003 by I. Groma, F. Csikor and M. Zaiser in a model taking into account the
 back stress describing more carefully  boundary layers (see
 \cite{GromaZai} for further details). The Groma-Balogh
 model neglects in particular the short range dislocation-dislocation
 correlations in one slip direction. For an extension to multiple slip
 see S. Yefimov and  E. Van der Giessen \cite[ch. 5.]{Yef1}. This multiple slip version of the Groma-Balogh model
 presents some analogies with some traffic flow
models (see O. Biham et al. \cite{Biham}). See also V. S. Deshpande et al. \cite{Des} 
for a similar model  with boundary conditions and exterior forces.  Recently, A. EL-Azab 
\cite{EL-Azab3D}, M. Zaiser, T. Hochrainer \cite{Zaiser3D} and R. Monneau
\cite{Regis3D} were interested in modeling the dynamics of dislocation
densities in the three-dimensional space, but many more  open questions
have to be  solved for establishing a satisfactory three-dimensional theory of
dislocations dynamics and for getting rigorous results.\\


\noindent  We stress out  the attention of the reader that there was no
  existence and uniqueness results for (\ref{EC:eq:i:1}). In
this paper we prove that (\ref{EC:eq:i:1}) admits a ``global in time''
solution.\\

\subsection{Main result}\label{EC:main}
In the present paper, we prove a ``global in  time'' existence result
for  the system (\ref{EC:eq:i:1}) describing the dynamics of
dislocation densities.

\noindent In this work, we consider the following
initial conditions:
\begin{equation}\tag{IC}\label{EC:initial}{\rho}^{\pm}(x_1,x_2,t=0)
={\rho}^{\pm}_0(x_1,x_2)=
{\rho}^{\pm,per}_0(x_1,x_2)+Lx_1,
\end{equation}
where ${\rho}^{\pm,per}$ is a $1$-periodic function in $x_1$ and
$x_2$. The periodicity is a way of studying the bulk behavior of
the material away from its boundary. Here $L$ is a  given positive
constant that represents the initial total dislocation densities of
$\pm$ type on the periodic cell.\\

\noindent Before to give our main result, we want to show  that the 
bilinear term on the right hand side of (\ref{EC:eq:i:1}) is well
defined. To this end, we need first to recall the following definition:

\begin{defi}\label{EC:defi:L1logL1}{\bf(The space  $L\log L$)}\\
 We define the space $L\log L(\T^2)$

$$L\log L(\T^2)=\left\{\mbox{$f\in L^1(\T^2)$ such that
  $\displaystyle{\int_{\T^2}|f|\ln\left(e+|f|\right)<+\infty}$}
 \right\}.$$

\noindent This space is endowed with the (Luxemburg) norm

$$\|f\|_{L\log
  L(\T^2)}=\inf\left\{\lambda>0:\displaystyle{\int_{\T^2}}\frac
  {|f|}{\lambda}\ln\left(e+\frac {|f|}{\lambda}\right)\le 1\right\}.$$

\end{defi}

\noindent The space $L\log L(\T^2)$ is a special space of Zygmund
 spaces (see  R. A. Adams \cite[(13), Page
234]{Adams}, E. M. Stein \cite[Page 43]{S93})

\noindent We can now state the following proposition.

\begin{pro}\label{EC:sens1}{\bf(Meaning of the bilinear term)}\\
Let $T>0$, $f$ and $g$ be two functions defined on  $\T^2\times(0,T)$,
such that \\$f\in L^1((0,T); W^{1,2}(\T^2))$ and $g\in L^{\infty}((0,T);
L\log L(\T^2))$ then,
 $$fg\in L^1(\T^2 \times (0,T)).$$
\end{pro}

\noindent We will see that the proof of this proposition (given in
Subsection \ref{EC:sens2}) is a direct
consequence  of Trudinger inequality.

\noindent We can now state our main result (see also our comments  in
Subsection \ref{com} on the
unknown uniqueness of the solution).

\begin{theo}\label{EC:theo:exi}{\bf (Global existence)}\\
 For all $T, L>0$ , and for every initial data
 $\rho_0^{\pm}\in L^{2}_{loc}(\R^2)$ with
\begin{enumerate}
\item[(H1)]
  $\rho_0^{\pm}(x_1+1,x_2)=\rho_0^{\pm}(x_1,x_2)+L,$ a.e. on $\R^2$,
\item[(H2)]
$\rho_0^{\pm}(x_1,x_2+1)=\rho_0^{\pm}(x_1,x_2),$ a.e. on $\R^2$,

\item[(H3)]
$\displaystyle{\frac{\partial{\rho_0}^{\pm}}{\partial x_1}}\geq 0,$
a.e. on $\R^2$, 
\item[(H4)]
$\left\|\displaystyle{\frac{\partial{\rho_0}^{\pm}}{\partial x_1}}\right\|_{L\log
L(\T^2)}\le C$, with  $\T^2=\R^2/\Z^2$,
\end{enumerate}

\noindent the system (\ref{EC:eq:i:1})-(\ref{EC:initial}) admits solutions $\rho^{\pm}\in
 C([0,T);L^1_{loc}(\R^2)) \cap L^{\infty}((0,T); L^2_{loc}(\R^2))$ in
 the distributional sense, such that,  $\rho^{\pm}(\cdot,t)$ satisfy
$(H1)$, $(H2)$, $(H3)$ and $(H4)$ for a.e. $t\in (0,T)$. Moreover, we have:
\begin{enumerate}
\item[(P1)]
$R_1^2R_2^2\left(\rho^+-\rho^-\right)\in L^2((0,T); W^{1,2}_{loc}(\R^2)).$
\end{enumerate}
\end{theo}
\begin{rem}{\bf (Bilinear term)}\\
It is clear here that  the bilinear term on the right hand side
of(\ref{EC:eq:i:1})  is always defined via $(P1)$ and
Proposition \ref{EC:sens1}.
\end{rem}

\noindent In order to prove our main  theorem we regularize the system (\ref{EC:eq:i:1}) by
 adding the viscosity term ($\e\Delta\rho^{\pm}$), and regularized also the initial data
(\ref{EC:initial}) by classical convolution. Then, using a fixed point
Theorem, we prove that our
regularized system admits local in time solutions. Moreover, as we get
some $\e$-independent {\it a priori} estimates we will be able  to extend our local in
time solution  into a global one. This turns out to be possible thanks to the
entropy inequality (\ref{EC:entropy}). Then, joined with other {\it a priori}
estimates, it will be possible to prove some compactness properties and
to pass to the limit  as $\e$ 
goes to $0$ is the $\e$-problem.\\

\begin{rem}{\bf (Entropy and energy inequalities)}\\
It turns out that the constructed solution also satisfies the following fundamental
entropy inequality (as a consequence of Lemma \ref{EC:lemme:entro}), for
a.e.  $t\in (0,T)$,

{\small \begin{equation}\begin{array}{ll}\label{EC:entropy}
\displaystyle{\int_{\T^2}}\sum_{\pm}\displaystyle{\frac{\partial{\rho}^{\pm}}{\partial
      x_1}}\ln\left(\displaystyle{\frac{\partial{\rho}^{\pm}}{\partial x_1}}\right)
+\displaystyle{\int_0^t}\int_{\T^2}\left(R_1R_2\left(\displaystyle{\frac{\partial{\rho}^{+}}{\partial
      x_1}}-
\displaystyle{\frac{\partial{\rho}^{-}}{\partial x_1}}\right)\right)^2\le 
\displaystyle{\int_{\T^2}\sum_{\pm}\displaystyle{\frac{\partial{\rho}^{\pm}_0}{\partial
      x_1}}\ln\left(\displaystyle{\frac{\partial{\rho}_0^{\pm}}{\partial x_1}}\right) }
\end{array}\end{equation}}

\noindent Moreover, (at least formally for sufficiently regular solution) the
following energy inequality holds:  
$$\begin{array}{ll}
\displaystyle{
\frac 12\int_{\T^2}\left(R_1R_2(\rho^+-\rho^-)(\cdot,t)\right)^2}
&+
\displaystyle{\int_0^t\int_{\T^2}\left(R_1^2R_2^2(\rho^+-\rho^-)\right)^2
\left(\displaystyle{\frac{\partial{\rho}^{+}}
{\partial x_1}}+\displaystyle{\frac{\partial{\rho}^{-}}{\partial
        x_1}}\right)}\le \\
&\displaystyle{
\frac 12\int_{\T^2}\left(R_1R_2(\rho^+_0-\rho^-_0)\right)^2}
\end{array}$$

\end{rem}

\begin{rem}{\bf (Bounds on the solution)}\\
If we denote $\rho=\rho^+-\rho^-$, then there exists a constant
$C$ independent on $T$, and a constant $C_T$ depending  on $T$ such that,\\

\noindent $(E1)$ $\|\rho^\pm-Lx_1\|_{L^{\infty}((0,T); L^2(\T^2))}\le C_T,$
$\;\;\;\;\;\;\;\;\;\;$ $(E4)$ $\|R_1^2R_2^2\rho\|_{L^2((0,T); W^{1,2}(\T^2))}\le C,$\\

\noindent $(E2)$ $\left\|\displaystyle{\frac{\partial{\rho}^{\pm}}{\partial x_1}}\right
\|_{L^{\infty}((0,T);L\log L(\T^2))}\le C,$ 
$\;\;\;\;\;\;\;\;\;\;\;\;\;\;$ $(E5)$  $\left\|\displaystyle{R_1^2R_2^2\frac{\partial{\rho}}{\partial t}}\right
\|_{L^{2}((0,T); W^{-1,2}(\T^2))}\le C,$\\\\

\noindent $(E3)$ $\left\|\displaystyle{\frac{\partial{\rho}^{\pm}}{\partial t}}\right
\|_{L^{2}((0,T); L^{1}(\T^2))}\le C,$
 \\\\

\noindent where $W^{-1,2}(\T^2)$ is the dual space of
$W^{1,2}(\T^2)$. 
\end{rem}

\noindent In a particular sub-case of model (\ref{EC:eq:i:1}) where the
dislocation densities depend on a single variable $x=x_1+x_2$, the existence and uniqueness of a
Lipschitz viscosity solution was proved in A. El Hajj, N. Forcadel
\cite{EF}. Also the existence
and uniqueness of a strong solution in $W^{1,2}_{loc}(\R\times
[0,T))$ was proved in  A. El Hajj \cite{EL}. Concerning the model of I. Groma, F. Csikor,
M. Zaiser \cite{GromaZai} which takes into consideration the short range
dislocation-dislocation  correlations giving a
parabolic-hyperbolic system, let us mention the work of H. Ibrahim
\cite{Ibrahim} where a result of existence and
uniqueness of a viscosity solution is given but only for a one-dimensional
model.\\

\noindent Our study of the dynamics of dislocation
densities in a special geometry is related to the more general 
dynamics of dislocation lines. We refer the interested reader to the
work of O. Alvarez et al.
\cite{AHLM04}, for a local
existence and uniqueness of some non-local Hamilton-Jacobi equation. We
also refer to O. Alvarez et al. \cite{ACM05} and  G.
Barles, O. Ley \cite{BL05}  for some long time existence results.
\subsection{Comments on the uniqueness of the solution and related literature}\label{com}

\noindent The problem (\ref{EC:eq:i:1}) is a system of transport 
equations with low regularity of the
vector field, so that the uniqueness of the solution here
is an open question. However, in the following 	we present some
uniqueness results where the vector field has a better regularity.\\

\noindent From a technical point of view, 
(\ref{EC:eq:i:1}) is related to other well known models, such
as  the transport equation with a low regularity  vector field. 
This equation was studied in the
work of  R. J. Diperna, P. L. Lions \cite{Dep} and L. Ambrosio
\cite{Amb2004}, where the authors showed the existence and uniqueness of
renormalized solutions by considering vector fields  in
$L^1((0,T);W^{1,1}_{loc}(\R^N))$ and $L^1((0,T); BV_{loc}(\R^N))$
respectively in both cases with bounded divergence. 
On the contrary in system (\ref{EC:eq:i:1}), we are only able to prove that for the
constructed solution, the vector field is in
$L^2((0,T);W^{1,2}_{loc}(\R^2))$  without any better estimate on the divergence
of the  vector field.\\

\noindent More generally  in the frame of symmetric hyperbolic systems, we refer to
the book of D. Serre \cite[Vol I, Th 3.6.1]{Serre12}, for a typical result of local
existence and uniqueness in   $C([0,T); H^s(\R^N))\cap C^1([0,T);
H^{s-1}(\R^N))$, with  $s> \frac N2+1$, by considering initial data
in  $H^s(\R^N)$. This result remains local in time, even in
dimension  $N=2$.\\ 


\noindent We can also remark that in the case where we multiply the
right side of the two equations in system (\ref{EC:eq:i:1}) by $-1$, we
get a quasi-geostrophic-like system. For those who are concerned in
quasi-geostrophic systems, we refer to P. Constantin et al.
\cite{Majda}, and to \cite{Majda1} for certain 2D numerical results. We also
refer 
to A. C\'ordoba, D. C\'ordoba
\cite{Cordoba},  D. Chae, A. C\'ordoba \cite{Cordoba1} for blow-up
results in finite time,  in dimension one.\\

\noindent Let us also mention some related Vlasov-Poisson models
(see J. Nieto et al.  \cite{Niet} for instance) and a related model in 
superconductivity studied by N. Masmoudi et al.  \cite{Mas} and by 
L. Ambrosio et al. \cite{Ambrosio}. These models were 
derived from some Vlasov-Poisson-Fokker-Planck 
models (see for instance T. Goudon et al. \cite{Gou} for an overview of similar
models). It is also worth mentioning that this model is related to 
Vlasov-Navier-Stokes equation see T. Goudon et
al. \cite{Vasseur1}, \cite{Vasseur2}.

\subsection{Notation}\label{EC:not}
In  what follows, we are going to use the following notation:
\begin{enumerate}
\item $\rho=\rho^{+}-\rho^{-}$,
\item $\rho^{\pm,per}(x_1,x_2,t)=\rho^{\pm}(x_1,x_2,t)-Lx_1$,
\item Let $f$ be a function defined on $\R^2\times(0,T)$ having values
  in $\R^2$, we denote by $f(t)=f(.,t): x\longmapsto f(x,t)$,
\item  Throughout the paper, $C$ is an arbitrary positive  constant
  independent on $T$ and $C_T$  is an arbitrary positive  constant
  depending on $T$.
\end{enumerate}

\subsection{Organization of the  paper }
 \noindent First, in Section 2, we recall the physical derivation of
system (\ref{EC:eq:i:1}).   In Section 3, we recall the definitions and
properties of some useful fundamental spaces, and 
we give the proof of Proposition
\ref{EC:sens1}.  We also prove that the bilinear term of our system has a better 
 mathematical meaning (see Proposition \ref{EC:est_bil}). Next, in Section 4, 
we regularize the initial conditions 
and  we show
that the system (\ref{EC:eq:i:1}), modified by a
term ($\e\displaystyle{\Delta \rho^{\pm}}$), admits local
solutions. Moreover,   we show that these solutions are regular
and increasing for all $t\in(0,T)$, for increasing initial data.
In Section 5, we prove some $\e$-uniform {\it a
  priori} estimates  for  the regularized solution obtained in Section
4. Then, thanks to these {\it a priori} estimates, we extend  the local in time
solutions for the $\e$-problem constructed in Section 4, in to global in
time solution.
Finally, in Section 6, we achieve the proof of our  main
theorem, passing to the limit in the equation as $\e$ goes to $0$, and
using some compactness properties inherited from our a priori estimates.
\section{Physical derivation of the model}\label{EC:subsce:model}
In this section we explain how to derive physically the system (\ref{EC:eq:i:1}). We
consider a three-dimensional crystal, with displacement  
$$u = (u_1,u_2,u_3):\R^3\rightarrow\R^3.$$ 
\noindent For $x=(x_1,x_2,x_3)$, and an
orthogonal basis $(e_1,e_2,e_3)$, we define the total strain by:

$$\displaystyle{\varepsilon_{ij}(u)= \frac{1}{2}}\left
(\displaystyle{\frac{\partial u_i}{\partial
    {x_j}}}+\displaystyle{\frac{\partial u_j}{\partial
    {x_i}}} \right),\;\;\; i,j=1,2,3.$$
\noindent This total strain is decomposed as 
$$\varepsilon_{ij}(u) = \e^e_{ij} + \e^p_{ij},$$

\noindent with $\e^e_{ij}$ is the elastic strain and $\e^p_{ij}$ the plastic strain
which is defined by:
\begin{equation}\label{EC:eq:epsilon0}
\varepsilon^p_{ij}=\rho\varepsilon^0_{ij},
\end{equation}
\noindent with the fixed matrix
$\displaystyle{\varepsilon^0_{ij}=\frac{1}{2}\left(1-\delta_{ij}\right)}$,
where $\delta_{ij}$ is the Kronecker symbol, in the special case of a single slip system where
dislocations move in the plane $\{x_2=0\}$ with Burgers vector
$\vec{b}=e_1$. Here $\gamma$ is
the resolved plastic strain, and will be clarified later. In the case of
linear homogeneous and isotropic elasticity, the stress
is given by
\begin{equation}\label{EC:homo:cond}\displaystyle{\sigma_{ij} = 2\mu\varepsilon^e_{ij}+
\lambda\delta_{ij}\left(\sum_{k=1,2,3}\varepsilon^e_{kk}\right)}\;\;\;\; \mbox{for $i,j=1,2,3$,}
\end{equation}

\noindent where $\lambda, \mu$ are the constant Lamé coefficients of the
crystal (satisfying $\mu >0$ and $3\lambda+2\mu>0$). Moreover the stress
satisfies the equation of elasticity:

$$\displaystyle{\sum_{j=1,2,3}\frac{\partial \sigma_{ij}}{\partial
    {x_j}}}=0.$$

\noindent We now assume that we are in a particular geometry where the
dislocations are straight lines parallel to the direction $e_3$ and that
the problem is invariant by translation in the $x_3$ direction. Moreover
we assume that $u_3=0$ and $\sigma_{i3}=0$ for $i=1, 2, 3$. Then, this
problem reduces to a two-dimensional
problem with $u_1,u_2$ only depending on $(x_1,x_2)$ and so we can
express  the resolved plastic  strain $\rho$ as 
$$\rho=\rho^+-\rho^-,$$
\noindent where $\displaystyle{\frac{\partial \rho^+}{\partial
    {x_1}}}$ and $\displaystyle{\frac{\partial \rho^-}{\partial
    {x_1}}}$ are respectively the densities of dislocations of Burgers
vectors  given by $\vec{b}=e_1$ and $\vec{b}=-e_1$.\\

\noindent Furthermore, these dislocation densities are transported in the direction
of  the Burgers  vector at a given velocity. This velocity is indeed the resolved shear  stress
$\displaystyle{\sum_{i,j=1,2,3}\sigma_{ij}\varepsilon^0_{ij}=\sigma_{12}}$, 
up to sign of the Burgers vectors. More
precisely, we have:
$$\displaystyle{\frac{\partial \rho^{\pm}}{\partial t}}=
\pm(\sigma_{12})\displaystyle{e_1.\nabla\rho^{\pm}}.$$

\noindent Finally, the functions $\rho^\pm$ and $u=(u_1,u_2)$ are  solutions of the coupled
system (see I. Groma, P. Balogh \cite{Groma}, \cite{Groma2}), on
$\R^2\times (0,T)$:

\begin{center}\begin{equation}\label{EC:coord_eq:elasdis}\left\{\begin{array} {lll}
\displaystyle{\sum_{j=1,2}\frac{\partial \sigma_{ij}}{\partial x_j}}  &
= 0& \mbox{for $i=1,2$,}\\
\sigma_{ij}&=  \displaystyle{2\mu\varepsilon^e_{ij}+
\lambda\delta_{ij}\left(\sum_{k=1,2}\varepsilon^e_{kk}\right)}&\mbox{for $i,j=1,2$,}\\
\varepsilon_{ij}^e &=\displaystyle{\frac{1}{2}}\left(\displaystyle{\frac{\partial u_i}{\partial
    {x_j}}}+\displaystyle{\frac{\partial u_j}{\partial
    {x_i}}} \right)-(\rho^{+}-{\rho}^{-})\varepsilon_{ij}^0
& \mbox{for $i,j=1,2$,}\\
\varepsilon_{ij}^0&=\displaystyle{\frac 12}\left(1-\delta_{ij}\right)&
  \mbox{for $i,j=1,2$,}\\
$\;$\\
\displaystyle{\frac{\partial \rho^{\pm}}{\partial t}} &
=\pm\sigma_{12}\displaystyle{\frac{\partial \rho^{\pm}}{\partial x_1}}.
 \end{array} \right.\end{equation}\end{center}

\noindent Then the following lemma holds.

\begin{lem}\label{EC:2D_1D}{\bf (Computation of $\sigma_{12}$)}\\
Assume that $(u_1,u_2)$ and $\rho=\rho^+-\rho^-$ are $\Z^2$-periodic
functions. If $(u_1,u_2)$, $\rho^+$, $\rho^-$ are solutions of 
problem (\ref{EC:coord_eq:elasdis}), then

\begin{equation}\label{EC:equi}
\sigma_{12}=-C_1\left(R_1^2R_2^2\rho\right),
\end{equation}

\noindent where  $\displaystyle{C_1=4\frac{(\lambda+\mu)\mu}{\lambda+2\mu}}>0$.
\end{lem}

\noindent Using this expression of $\sigma_{12}$ and rescaling in time
with the positive constant $C_1$ we obtain system  (\ref{EC:eq:i:1}),
from the last equation (\ref{EC:coord_eq:elasdis}).

\noindent {\bf Proof of Lemma \ref{EC:2D_1D}:}\\
\noindent We can rewrite the first equation of
(\ref{EC:coord_eq:elasdis}) with $\div\; u = \displaystyle{\frac{\partial u_1}{\partial
    {x_1}}}+\displaystyle{\frac{\partial u_2}{\partial
    {x_2}}}$

\begin{subequations}\label{EC:equi1}
\begin{gather}
\mu\Delta u_1+ (\lambda+\mu)\displaystyle{\frac{\partial }{\partial
    {x_1}}} (\div\; u)= \mu \displaystyle{\frac{\partial \rho}{\partial
    {x_2}}},\label{EC:equi11}\\
\mu\Delta u_2+ (\lambda+\mu)\displaystyle{\frac{\partial }{\partial
    {x_2}}} (\div\; u)= \mu \displaystyle{\frac{\partial \rho}{\partial
    {x_1}}}.\label{EC:equi12}
\end{gather}\end{subequations}

\noindent Considering $\displaystyle{\frac{\partial }{\partial
    {x_1}}}$(\ref{EC:equi11})+$\displaystyle{\frac{\partial }{\partial
    {x_2}}}$(\ref{EC:equi12}), we get

$$\displaystyle{(\lambda+2\mu)\Delta(\div\; u)=2\mu\frac{\partial^2 \rho}{\partial
    {x_1}\partial {x_2}}}.$$

\noindent Plugging the expression of $\div\; u$ into (\ref{EC:equi1}), we get

\begin{subequations}\label{EC:equi2}
\begin{gather}
\Delta u_1= \displaystyle{\frac{\partial \rho}{\partial
    {x_2}}}-2\frac{(\lambda+\mu)}{(\lambda+2\mu)}\displaystyle{\frac{\partial }{\partial
    {x_1}}}\Delta^{-1} \displaystyle{\frac{\partial^2 \rho}{\partial {x_1}\partial {x_2}}
  },\label{EC:equi21}\\
\Delta u_2= \displaystyle{\frac{\partial \rho}{\partial
    {x_1}}}-2\frac{(\lambda+\mu)}{(\lambda+2\mu)}\displaystyle{\frac{\partial }{\partial
    {x_2}}}\Delta^{-1} \displaystyle{\frac{\partial^2 \rho}{\partial {x_1}\partial {x_2}}
  }.\label{EC:equi22}
\end{gather}\end{subequations}

\noindent Considering now $\displaystyle{\frac{\partial }{\partial
    {x_2}}}$(\ref{EC:equi21})+$\displaystyle{\frac{\partial }{\partial
    {x_1}}}$(\ref{EC:equi22}) , we obtain

\begin{equation}\label{EC:eq:force}\Delta \left(\displaystyle{\frac{\partial u_1}{\partial
    {x_2}}}+\displaystyle{\frac{\partial u_2}{\partial
    {x_1}}}\right)=\Delta
(\rho^{+}-\rho^{-})-4\frac{(\lambda+\mu)}{(\lambda+2\mu)} \Delta^{-1}\displaystyle{\frac{\partial^4}{\partial
          {x_1^2}\partial{x_2^2}}(\rho^{+}-\rho^{-})}.\end{equation}

\noindent Recalling that

\begin{equation}\label{EC:eq:sigma1}\sigma_{12}=\mu
\left(\displaystyle{ \left(\displaystyle{\frac{\partial u_1}{\partial
    {x_2}}}+\displaystyle{\frac{\partial u_2}{\partial
    {x_1}}}\right)-( \rho^{+}-\rho^{-})}\right),
\end{equation}

\noindent  this yields $\sigma_{12}=-4\frac{(\lambda+\mu)\mu}{(\lambda+2\mu)}
\Delta^{-2}\displaystyle{\frac{\partial^4}{\partial
          {x_1^2}\partial{x_2^2}}(\rho^{+}-\rho^{-})}=
        -C_1\left(R_1^2R_2^2(\rho^{+}-\rho^{-})\right).$  $\hfill\Box$

\begin{rem}{\bf (Property  of the elastic energy)}\\
If we define the elastic energy by 
$$\displaystyle{E=
\int_{\R^2/\Z^2}\mu\sum_{i,j=1,2}(\e_{ij}^e)^2+\frac{\lambda}{2}\left(\sum_{k=1,2}\e_{kk}^e\right)^2.
}$$ 

\noindent Using system (\ref{EC:coord_eq:elasdis}) we can show formally
that 

$$\displaystyle{\frac{d E }
{dt }}
=-\int_{\R^2/\Z^2}(\sigma_{12})^2\left(\displaystyle{\frac{\partial \rho^+}{\partial
    {x_1}}}+ \displaystyle{\frac{\partial \rho^-}{\partial
    {x_1}}}\right)\le 0.
$$
\noindent where we have used the fact that $\displaystyle{\frac{\partial \rho^+}{\partial
    {x_1}}},\displaystyle{\frac{\partial \rho^-}{\partial
    {x_1}}}\ge 0$ to see that the elastic energy
is a non-increasing in time.  Hence, the  elastic energy $E$ is a Lyapunov functional for our
dissipative model.
\end{rem}


\section{Concerning the meaning of the solution of (\ref{EC:eq:i:1})}\label{EC:sens}
In this section we prove Proposition \ref{EC:sens1}. This shows that if
(\ref{EC:eq:i:1}) admits solutions verifying the conditions of
Theorem \ref{EC:theo:exi}, then we can give a mathematical meaning to the
bilinear term. In order to do this, we need to define some
functional spaces and recall some of their 
properties, that will be  used later in our work.

\subsection{Properties of  some useful Orlicz spaces}\label{EC:Orli}
We recall the definition of  Orlicz spaces and 
some of their properties. For details, we refer to R. A. Adams \cite[Ch. 8]{Adams} and
M. M. Rao, Z. D. Ren \cite{Rao}.

\noindent A real valued function $\displaystyle{A
  :[0,+\infty)\rightarrow \R}$ is called a Young
function  if  it has the following properties (see R. O'Neil \cite[Def 1.1]{Neil}):
\begin{itemize}
\item  $A$ is a continuous, non-negative, non-decreasing and convex function.
\item $A(0)=0$ and  $\displaystyle{\lim_{t\rightarrow
    +\infty}A(t)=+\infty}$.
\end{itemize}
\noindent Let $A(\cdot)$ be a Young function. The Orlicz class $K_A(\T^2)$ is the
set of (equivalence classes of) real-valued measurable function $h$ on
$\T^2$ satisfying

$$\displaystyle{\int_{\T^2} A(|h(x)|)< +\infty}.$$

\noindent The Orlicz space $L_A(\T^2)$ is the linear hull of
$K_A(\T^2)$ supplemented with the Luxemburg norm

$$\|f\|_{L_A(\T^2)}=\inf\left\{\lambda>0:\displaystyle{\int_{\T^2}}
A\left(\frac {|h(x)|}{\lambda}\right)\le 1\right\}.$$

\noindent Endowed with this norm, the Orlicz space $L_A(\T^2)$ is a Banach
space. Moreover, for all $f\in L_A(\T^2)$, we have the following estimate
\begin{equation}\label{norm}\|f\|_{L_A(\T^2)}\le 1+\int_{\T^2}A(|f(x)|)\end{equation}

\begin{defi}{\bf(Some Orlicz spaces)}\\

\noindent $\bullet\;\;\mbox{$EXP_{\a}(\T^2)$ denotes the Orlicz space
defined by the function $A(t)=e^{t^\a}-1$ for $\a \ge 1$}.$\\

\noindent $\bullet\;\;\mbox{$L\log^{\beta}L(\T^2)$ denotes the Orlicz space
defined by the function
$A(t)=t(\log(e+t))^{\beta}$, for $\beta\ge 0$}.$

\end{defi}
\noindent Observe that
for $0<\beta\le 1$  the  space $EXP_{\frac 1\beta}(\T^2)$  is the dual of
the Zygmund space $L\log^{\beta}L(\T^2)$. (see C. Bennett and
R. Sharpley \cite[Def 6.11]{BE88}). It is worth noticing that $L
\log^1L(\T^2)= L \log L(\T^2)$.\\

\noindent Let us recall some useful properties of these spaces. The first one is the 
generalized Hölder inequality.

\begin{lem}\label{EC:hold}{\bf(Generalized Hölder inequality)}\\
\noindent i) Let $f\in EXP_2(\T^2)$ and $g\in L\log^{\frac 12} L(\T^2)$.
Then there exists a constant $C$ such that (see R. O'Neil \cite[Th 2.3]{Neil})
$$\|fg\|_{L^1(\T^2)}\le C \|f\|_{EXP_2(\T^2)}\|g\|_{L\log^{\frac 12}
  L(\T^2)}.$$

\noindent ii) Let $f\in EXP_2(\T^2)$ and $g\in L\log L(\T^2)$. Then 
there exists a constant $C$ such that (see R. O'Neil \cite[Th 2.3]{Neil})
$$\|fg\|_{L\log^{\frac 12} L(\T^2)}\le C\|f\|_{EXP_2(\T^2)}\|g\|_{L\log
  L(\T^2)}.$$
\end{lem}
\noindent The second property is the Trudinger inequality.
 
\begin{lem}\label{EC:trud}{\bf(Trudinger inequality)}\\
 There exists a constant $\gamma>0$ such that, for all $f\in W^{1,2}(\T^2)$,
 we have (see N. S. Trudinger \cite{Trud})
 
$$\displaystyle{\int_{\T^2}\displaystyle{e}^{\displaystyle{\gamma}\left(\displaystyle{\frac
      {f}{\|f\|_{W^{1,2}(\T^2)}}}\right)^2}}\le 1.$$

\noindent In particular we have the following embedding
$$W^{1,2}(\T^2)\hookrightarrow EXP_2(\T^2).$$

\end{lem}




\subsection{Sharp estimate of the bilinear term}\label{EC:sens2}
\noindent Now, we propose to verify with the help of the following
proposition that the system  (\ref{EC:eq:i:1}) has indeed a sense, and
first prove a better estimate than those mentioned in Proposition \ref{EC:sens1}. Namely, we
have the following.

\begin{pro}\label{EC:est_bil}{\bf(Estimate of the bilinear term)}\\
Let $T>0$, $f$ and $g$ be two functions defined on  $\T^2\times(0,T)$,
such that
\begin{enumerate}
\item[(1)] $f\in L^2((0,T); W^{1,2}(\T^2))$,
\item[(2)] $g\in L^{\infty}((0,T); L\log L(\T^2))$. Then
 \end{enumerate}
$$fg\in L^{2}((0,T); L\log^{\frac 12} L(\T^2))$$

\noindent and for a positive constant $C$, we have:

$$\|fg\|_{L^{2}((0,T); L\log^{\frac 12} L(\T^2))}\le 
C\|f\|_{L^2((0,T); W^{1,2}(\T^2))}\|g\|_{L^{\infty}((0,T); L\log
  L(\T^2))}.$$

\end{pro}

\noindent For the proof of this Proposition ,
 we use Lemma \ref{EC:hold} (ii), and  integrate
in time. Thanks to the  Trudinger inequality (Lemma \ref{EC:trud}), we
get the result. We do the same way for the proof of the Proposition
\ref{EC:sens1}.

\section{Local existence of solutions of a regularized system}
In this section, we state a local in time
existence result for system (\ref{EC:eq:i:1}), modified by the term $\e\Delta
\rho^{\pm}$, and for smoothed data. This modification
brings us to study, for all $0<\e\le1$, the following regularized system:
\begin{equation}\tag{$P_\e$}\label{EC:eq:i:2}\left\{\begin{array} {lll}
\displaystyle{\frac{\partial{\rho}^{+,\varepsilon}}{\partial t}}-\varepsilon\Delta{\rho}^{+,\varepsilon}
& =-(R_1^2R_2^2\rho^\varepsilon)\displaystyle{\frac{\partial{\rho}^{+,\varepsilon}}{\partial
    x_1}}
&\mbox{in  $\mathbb \D'(\R^2\times(0,T))$,} \\
 \\
\displaystyle{\frac{\partial{\rho}^{-,\varepsilon}}{\partial
    t}}-\varepsilon\Delta{\rho}^{-,\varepsilon}
& =\;\;\;(R_1^2R_2^2\rho^\varepsilon)\displaystyle{\frac{\partial{\rho}^{-,\varepsilon}}{\partial
    x_1}}
&\mbox{in  $\mathbb \D'(\R^2\times(0,T))$,}\\
\end{array} \right.\end{equation}
where $\rho^\varepsilon=\rho^{+,\varepsilon}-\rho^{-,\varepsilon}$, with
the following regular initial data:

\begin{equation}\tag{$IC_{\e}$}\label{EC:initialapp}\rho^{\pm,\e}(x,0)=\rho_0^{\pm,\e}(x)=
\rho^{\pm,per}_0\ast\eta_{\e}(x)+(L+\e)x_1=\rho^{\pm,\e,per}_0(x)+L_{\e}x_1,\end{equation}
where $\eta_{\e}(\cdot)=\frac
{1}{\e^2}\eta(\frac{\cdot}{\e})$, such that $\eta\in
C^{\infty}_c(\R^2)$ is a non-negative function and  $\int_{\R^2}\eta=1$.

\begin{rem}{$\;$}\\
We consider  $L_{\e}=L+\e$  to obtain strictly monotonous  initial data
$\rho_0^{\pm,\e}$. 
This condition will be useful in the proof of  Lemma \ref{EC:lemme:entro}.
\end{rem}

\noindent For  the regularized system
(\ref{EC:eq:i:2})-(\ref{EC:initialapp}) we have the following result. 

\begin{theo}\label{EC:theo:exip_regu}{\bf (Local existence result of monotone
    smooth solutions)}\\
For all initial data $\rho^{\pm}_0\in L^{2}_{loc}(\R^2)$
 satisfying  $(H1)$, $(H2)$ and $(H3)$, and all $\e>0$, there exists $T^\star>0$
such that the system (\ref{EC:eq:i:2})-(\ref{EC:initialapp}) admits
solutions $\rho^{\pm,\e}\in C^{\infty}(\R^2\times [0,T^{\star}))$.  Moreover $\rho^{\pm,\e}(\cdot,t)$ satisfy 
 $(H1)$, $(H2)$ and
 $\displaystyle{\frac{\partial
    \rho^{\pm,\e}}{\partial x_1}>0}$,  for all $t\in [0,T^{\star})$.
\end{theo}

\noindent Before proving  Theorem \ref{EC:theo:exip_regu}, let us recall some
well known results. \\

\noindent We first  recall the Picard fixed point result which will be
applied in the proof of this theorem  in order to prove, the existence of solutions.

\begin{lem}\label{EC:lem:pointfixe}{\bf (Picard Fixed point Theorem)}\\
Let $E$ be a Banach space, $B$ is a continuous bilinear application
over $E\times E$ having values in $E$, and  $A$ a continuous linear
application over $E$ having values in $E$ such that:\\
$$\|B(x,y)\|_{E}\leq\eta\|x\|_{E}\|y\|_{E}\;\;\;\mbox{for all}\;\;\;x,y\in E,$$
$$\|A(x)\|_{E}\leq\mu\|x\|_{E}\;\;\;\mbox{for all}\;\;\;x\in E, $$
where $\eta>0$ and $\mu\in(0,1)$ are two given constants. Then, for
every $x_0\in E $ verifying
$$\|x_0\|_{E}<\frac{1}{4\eta}(1-\mu)^2,$$
the equation $x=x_0+B(x,x)+A(x)$ admits a solution in $E$.
\end{lem}
For the proof of Lemma \ref{EC:lem:pointfixe}, see M. Cannone
\cite[Lemma 4.2.14]{CANN95}.\\

\noindent  We now  recall the  following decay estimates  for the heat semi-group.

\begin{lem}\label{EC:estsemi}{\bf(Decay estimate)}\\
Let $r,p,q\ge 1$. Then, for all functions  $f\in L^{q}(\T^2)$ and  $g\in L^p(\T^2)$,
where $\displaystyle{\frac 1r\le \frac 1q+\frac 1p}$, we have, for
$S_{1}(t)=e^{t\Delta}$, the following estimates:\\

$i)\;\;\; \|S_{1}(t)(fg)\|_{L^{r}(\T^2)}\leq C t^{-\left(\frac 1p+\frac
  1q-\frac 1r\right)}\|f\|_{L^q(\T^2)}\|g\|_{L^{p}(\T^2)}\;\; \mbox{for all $t>0$,} $\\

$ii)\;\;\;\|\nabla S_{1}(t)(fg)\|_{L^{r}(\T^2)}\leq C t^{-\left(\frac 12+\frac 1p+\frac
  1q-\frac 1r\right)}\|f\|_{L^q(\T^2)}\|g\|_{L^{p}(\T^2)}\;\; \mbox{for all $t>0$,}$\\

\noindent where   $C$ is a positive constant depending only on $r, p, q$.
\end{lem}

\noindent  The proof of this lemma is a direct application  of  the classical version of the  $L^r$-$L^p$ estimates for the heat
semi-group (see A. Pazy \cite[Lemma 1.1.8, Th 6.4.5]{Pazy}) and the  Hölder inequality.

\noindent Here is now, the demonstration of Theorem \ref{EC:theo:exip_regu}. \\

\noindent {\bf Proof of  Theorem \ref{EC:theo:exip_regu}:} \\
 \noindent  Frist we prove using Lemma  \ref{EC:lem:pointfixe} the local existence of  
the regularized system (\ref{EC:eq:i:2})-(\ref{EC:initialapp}).  This result
is achieved in a super-critical space.  Here particularly we  chose
the space of functions $L^{\infty}((0,T);W^{1,\frac 32}_{loc}(\R^2))$.  The notation "super-critical space" is to say that we are choosing a space where our 
$\e$-problem is well defined,  and where the right hand term (the bilinear term) is in a space better than $L^1$. This premits to use a bootstrap
 arguments  which easily leads to the existence of smooth solution of the regularized problem.

 \noindent Now, we note  that,  if we let  $\rho^{\pm,\e,per}=\rho^{\pm,\e}-L_{\e}x_1$, we know
that the system (\ref{EC:eq:i:2}) is equivalent to,
\begin{equation}\tag{$P_\e^{per}$}\label{EC:eq:i:3}\displaystyle{\frac{\partial{\rho}^{\pm,\varepsilon}}{\partial t}}
-\varepsilon\Delta\rho^{\pm,\varepsilon,per}
=\mp(R_1^2R_2^2\rho^{\varepsilon})
\displaystyle{\frac{\partial{\rho}}{\partial
      x_1}^{\pm,\varepsilon,per}}\mp
  L_{\e}(R_1^2R_2^2\rho^{\varepsilon})\;\;\mbox{in $\D'(\T^2\times(0,T))$,}\end{equation}
with  initial conditions,
\begin{equation}\tag{$IC_{\e}^{per}$}\label{EC:initialapp1}\rho^{\pm,\e,per}(x,0)
=\rho_0^{\pm,\e}(x)-L_{\e}x_1=
\rho_0^{\pm,\e,per}(x).\end{equation}

\noindent  To solve this  system  in the space $L^{\infty}((0,T);W^{1,\frac 32}(\T^2))$ we reduce to construct a solution $\rho^{\pm,\varepsilon,per}$ 
to the following integral problem (see A. Pazy \cite[Th 5.2, Page 146]{Pazy})

\begin{equation}\begin{array}{ll}\tag{$In_{\e}$}\label{EC:eq:i:6}
\displaystyle{\rho^{\pm,\varepsilon,per}(\cdot,t)}=S_{\e}(t)\rho^{\pm,\e,per}_0
&\mp L_{\e}\displaystyle{\int_0^t}
S_{\e}(t-s)\left(R^2_1R^2_2\rho^{\varepsilon}(s)\right)ds\\
\\
&\mp\displaystyle{\int_0^t}
S_{\e}(t-s)\left(\left(R^2_1R^2_2\rho^{\varepsilon}(s)\right)
\displaystyle{\frac{\partial{\rho}}{\partial
    x_1}^{\pm,\varepsilon,per}(s)}\right)ds,\end{array}\end{equation}

 \noindent where $S_{\e}(t)= S_1(\e t)$, and $S_1(t)=e^{t\Delta }$ is a the heat semi-group.  We rewrite the system (\ref{EC:eq:i:6}) in t
 he following vectorial  form:

{\small $$\displaystyle{\rho_v^{\e}(x,t)}=S_{\e}(t)\rho_{0,v}^{\e}
+L_{\e}\bar{J_1}\displaystyle{\int_0^t} S_{\e}(t-s)\left(R^2_1R^2_2\rho^{\varepsilon}(s)\right)ds
+\bar{I_1}\displaystyle{\int_0^t}S_{\e}(t-s)\left(R^2_1R^2_2\rho^{\varepsilon}(s)\right)\displaystyle
{\frac{\partial{\rho_v^{\varepsilon}}}{\partial
    x_1}(s)}ds,$$}

\noindent where $S_{\e}(t)=S_{1}(\e t)$,  {\small $\rho_v^{\e}=(\rho^{+,\varepsilon,per},\rho^{-,\varepsilon,per})$},
{\small $\rho_{0,v}^{\e}=(\rho^{+,\e,per}_0,\rho^{-,\e,per}_0)$},
{\small $\bar{I_1}=\left(\begin{array}{cc}
                            -1 & 0 \\
                          0 & 1
        \end{array}\right)$ } and {\small $\bar{J_1}=\left(\begin{array}{c}
                            -1 \\
                           1
                        \end{array}\right)$}.\\

\noindent Which is equivalent to,

\begin{equation}\label{EC:eq:point}\displaystyle{\rho_v^{\varepsilon}(x,t)}=S_{\e}(t)\rho_{0,v}^{\e}+
B(\rho_v^{\varepsilon},\rho_v^{\varepsilon})(t)+A(\rho_v^{\varepsilon})(t),\end{equation}

\noindent where $B$ is a bilinear map and $A$ is a linear
one defined respectively, for every vector $u=(u_1,u_2)$ and
$v=(v_1,v_2)$, as follows:
\begin{equation}\label{EC:de:bil}\displaystyle{B(u,v)(t)}=\bar{I_1}\int_0^t
S_{\e}(t-s)\left(\left(R^2_1R^2_2(u_1-u_2)\right)\displaystyle{\frac{\partial{v}}{\partial
    x_1}(s)}\right)ds,\end{equation}

\begin{equation}\label{EC:de:li}\displaystyle{A(u)(t)=L_{\e}\bar{J_1}\int_0^t
  S_{\e}(t-s)\left(R^2_1R^2_2(u_1-u_2)(s)\right)}ds.\end{equation}\\
\noindent Now, we  apply Lemma \ref{EC:lem:pointfixe} to
equation  (\ref{EC:eq:point}). First of all, we estimate the bilinear term,\\
$$\begin{array}{ll}\|B(u,v)(t)\|_{(W^{1,\frac 32}(\T^2))^2}
&\leq\left\|\bar{I_1}\displaystyle{\int_0^t}
S_{\e}(t-s)\left(\left(R^2_1R^2_2(u_1-u_2)\right)\displaystyle{\frac{\partial{v}}{\partial
    x_1}(s)}\right)ds\right\|_{(W^{1,\frac 32}(\T^2))^2}\\
\\
&\leq
\displaystyle{\int_0^t}\left\|
S_{\e}(t-s)\left(\left(R^2_1R^2_2(u_1-u_2)\right)\displaystyle{\frac{\partial{v}}{\partial
    x_1}(s)}\right)ds\right\|_{(W^{1,\frac 32}(\T^2))^2}.
\end{array}$$
\noindent Then, since $L^4(\T^2)\hookrightarrow L^{\frac 32}(\T^2)$, we
have,

\begin{equation}\begin{array}{ll}\label{EC:est33}
\hspace{-0.5cm}\|B(u,v)(t)\|_{(W^{1,\frac 32}(\T^2))^2}
&\hspace{-0.3cm}\leq\displaystyle{\int_0^t}\left\|
S_{\e}(t-s)\left(\left(R^2_1R^2_2(u_1-u_2)\right)\displaystyle{\frac{\partial{v}}{\partial
    x_1}(s)}\right)ds\right\|_{(L^{4}(\T^2))^2}\\
\\
&\hspace{-0.3cm}+
\displaystyle{\int_0^t}\left\|\nabla
S_{\e}(t-s)\left(\left(R^2_1R^2_2(u_1-u_2)\right)\displaystyle{\frac{\partial{v}}{\partial
    x_1}(s)}\right)ds\right\|_{(L^{\frac 32}(\T^2))^2}.
\end{array}\end{equation}

\noindent We use Lemma \ref{EC:estsemi} (i) with  $r=4,q=3,p=\frac 32$ 
to estimate the first term and Lemma \ref{EC:estsemi} (ii) with $r=\frac
32,q=4,p=\frac 32$ to estimate the second term. We get for $0\le t\le
T$, and with constants $C$ depending on $\e$,

$$\begin{array}{ll}\|B(u,v)(t)\|_{(W^{1,\frac 32}(\T^2))^2}

&\hspace{-0.3cm}\leq C\displaystyle{\int_0^t\frac{1}{(t-s)^{\frac{3}{4}}}}\left\|R^2_1R^2_2u(s)\right\|_{(L^4(\T^2))^2}
\left\|\displaystyle{\frac{\partial{v}}{\partial x_1}(s)}\right\|_{(L^{\frac 32}(\T^2))^2}ds\\
\\
&\hspace{-0.3cm}\leq C\displaystyle{\sup_{0\leq s<T}}(\left\|u(s)\right\|_{(W^{1,\frac 32}(\T^2))^2})
\sup_{0\leq
  s<T}(\left\|v(s)\right\|_{(W^{1,\frac 32}(\T^2))^2})\displaystyle{\int_0^t\frac{1}{(t-s)^{\frac{3}{4}}}}ds.
\end{array}$$
 
\noindent Here we have used in the second line  the property that Riesz
transformations are continuous from $L^{\frac 32}$ onto itself (see A. Zygmund \cite[Vol I, Page 254, (2.6)]{ZY59}) 
and the Sobolev
injection $W^{1,\frac 32}(\T^2)\hookrightarrow L^4(\T^2)$. Hence we have,
\begin{equation}\label{EC:eq:bi}\|B(u,v)\|_{L^{\infty}((0,T);(W^{1,\frac 32}(\T^2))^2)}\leq
\eta(T)\|u\|_{L^{\infty}((0,T);(W^{1,\frac 32}(\T^2))^2)}\|v\|_{L^{\infty}((0,T);(W^{1,\frac 32}(\T^2))^2)},
\end{equation}
\noindent  with $\eta(T)=C_0T^{\frac{1}{4}}$ for some constant $C_0>0$.
We estimate the linear term in the same way to get,
\begin{equation}\label{EC:eq:li}\|A(u)\|_{L^{\infty}((0,T);(W^{1,\frac
      32}(\T^2))^2)}\leq L_{\e}\eta(T)
\|u\|_{L^{\infty}((0,T);(W^{1,\frac 32}(\T^2))^2)}.
\end{equation}
\noindent Moreover, we know by classical properties of heat semi-group that,
\begin{equation}\label{EC:eq:do}\| S_{\e}(t)\rho_{0,v}^{\e}\|_{L^{\infty}((0,T); (W^{1,\frac
    32}(\T^2))^2)}\le \| \rho_{0,v}^{\e}\|_{(W^{1,\frac
    32}(\T^2))^2}.\end{equation}
\noindent Now, if we take
 \begin{equation}\label{EC:tstar}(T^{\star})^{\frac
     14}=\min\left(\frac{1}{2C_0L_{\e}},\frac{1}{16C_0\|\rho_{0,v}^{\e}\|_{(W^{1,\frac
32}(\T^2))^2}}\right),\end{equation}

\noindent we can 
easily verify that we have the following inequalities:

\begin{equation}\label{EC:tet}\|\rho_{0,v}^{\e}\|_{(W^{1,\frac 32})^2(\T^2)}<
\frac{1}{4\eta(T^{\star})}(1-L_{\e}\eta(T^{\star}))^2, \;\;\; \mbox{and
  $L_{\e}\eta(T^{\star})<1$,}\end{equation}

\noindent Using inequalities (\ref{EC:eq:bi}), (\ref{EC:eq:li}),
(\ref{EC:eq:do}), (\ref{EC:tet}) and Lemma \ref{EC:lem:pointfixe} with the
space \\$E=\left(L^{\infty}((0,T^{\star}); W^{1,\frac
32}(\T^2))\right)^2$, we show the local in time  existence  
or the system (\ref{EC:eq:point}) in $ \left(L^{\infty}((0,T^{\star}); W^{1,\frac
32}(\T^2))\right)^2$.   As a consequence we prove that  the system
(\ref{EC:eq:i:2})-(\ref{EC:initialapp}) admits some solutions   $\rho^{\pm,\e}\in L^{\infty}((0,T^{\star});
W^{1,\frac 32}_{loc}(\R^2))$, satisfying $(H1)$ and $(H2)$ a.e. $t\in
[0,T^{\star})$. \\

\noindent  Finally,   the fact that  product  $\left(R^2_1R^2_2\rho^{\varepsilon}\right)
\displaystyle{\frac{\partial{\rho}}{\partial
    x_1}^{\pm,\varepsilon,per}}$ is well defined in $L^{\infty}((0,T);L^{\frac 65}(\T^2))$ since
$L^{\infty}((0,T);W^{1,\frac 32}(\T^2))\hookrightarrow L^{\infty}((0,T);L^6(\T^2))$,  we can  prove,  by   a  
 bootstrap argument,  the regularity of the solution.  The monotonicity of the solution is a consequence of the  
 maximum principle  for scalar parabolic equations the previous result  (see G. Lieberman \cite[Th 2.10]{Lieb}). 

$\hfill\Box$




\section{$\e$-Uniform estimates on the solution of the regularized
  system}
In this section, we  prove some fundamental $\e$-uniform estimates. In 
Subsection \ref{EC:uslel} we give some general estimates which are 
independent  on the equation. In the second Subsection
\ref{EC:apriori} we establish a priori estimates on
the solutions of system (\ref{EC:eq:i:2}).


\subsection{Useful estimates}\label{EC:uslel}

Now we recall some well known properties of Riesz transform, that will
be used later in our work.

\begin{lem}{\bf(Properties of Riesz transform)}\label{Riesz}\\
\noindent  i)  For all $g\in L^p(\T^2)$, $1<p<+\infty$, we have  
$$\|R_ig\|_{L^p(\T^2)}\le \|g\|_{L^p(\T^2)}.$$

\noindent ii) If $g\in L^2(\T^2)$, then
$\displaystyle{\int_{\R/\Z}R_1g(x_1,x_2)dx_1=0}$, for a.e. $x_2\in
\R/\Z$.

\noindent iii) For all $g\in L^2(\T^2)$, we have
$\displaystyle{\frac{\partial }{\partial x_1}}R_2 g
=\displaystyle{\frac{\partial }{\partial x_2}}R_1 g$ and  
$R_1R_2g=R_2R_1g$.

\noindent iv) For all $f, g\in L^2(\T^2)$, we have $\displaystyle{\int_{\T^2}(R_if)g=\int_{\T^2}f(R_ig).}$

\noindent v)  If
$g\in L^2(\T^2)$ and does not depend on $x_2$, then $R_1 g =0$.

\end{lem}
\noindent{\bf Proof of Lemma \ref{Riesz}:}\\
For the proof of i) (see A. Zygmund \cite[Vol I, Page 254, (2.6)]{ZY59}).  The proof of  iv) this is straightforward, using Fourier series. 
For the  proof of ii), it suffices to note that,  
if we denote by
$f(x_2)=\displaystyle{\int_{\R/\Z}R_1g(x_1,x_2)dx_1}$, then we have
 $c_{k_2}(f)= c_{(0,k_2)}(R_1g)=0$  by definition of $c_k$ for $k_1=0$. Finally, we prove  iii), checking simply that 

$$\begin{array}{ll}\displaystyle{c_k\left(\displaystyle{\frac{\partial }{\partial x_1}}R_2 g\right)}
= \displaystyle{2\pi i k_1 \frac {k_2}{|k|}
c_k(g)}= \displaystyle{2\pi i k_2 \frac {k_1}{|k|}
c_k(g)}
=\displaystyle{c_k\left(\displaystyle{\frac{\partial }{\partial x_2}}R_1 g\right)},
\end{array}$$

\noindent and similar we prove second equality of iii).  The  prove of v) is direct.  In fact,

$$\displaystyle{c_{(k_1,k_2)}(R_1g)=\frac {k_1}{|k|}\int_{\T^2}g(x_2)e^{-2\pi
    i (k_1x_1+k_2x_2)}dx_1dx_2=0}.$$

$\hfill\Box$

\begin{lem}{\bf($L^{\infty}$ estimate)}\label{EC:lem:ebmo}\\
If $f\in L^2_{loc}(\R^2)$ and $f$ 
 verifies
$(H1)$, $(H2)$ and $(H3)$ for a.e. $t\in (0,T)$, then there exists a
constant  $C=C(L)$ such that,

\begin{equation}\label{EC:moy}\left\|f^{per}-\displaystyle{\int_0^1f^{per} dx_1}\right\|_{
L^{\infty}(\T^2)}\le C.\end{equation}

\noindent where $f^{per}=f-Lx_1$.
\end{lem}
\noindent{\bf Proof of Lemma \ref{EC:lem:ebmo}:}

\noindent We compute 

$$\begin{array}{ll}\displaystyle{\int_0^1\left|\frac{\partial f^{per}}{\partial
      x_1}\right|dx_1}
=\displaystyle{\int_0^1\left|\frac{\partial f}{\partial
      x_1}-L\right|dx_1}
&\le L+ \displaystyle{\int_0^1\left|\frac{\partial f}{\partial
      x_1}\right|dx_1} \\
&= L+ \displaystyle{\int_0^1\frac{\partial f}{\partial
      x_1}dx_1}    \\
&= 2L,
\end{array}$$

\noindent  where  we use $(H3)$ in the second line and $(H1)$ in the
last line. We next apply a ``Poincaré-Wirtinger inequality'' in $x_1$ and  we deduce
the result. $\hfill\Box$

\noindent We will also use the following technical result.

\begin{lem}{\bf($L\log L$ Estimate)}\label{EC:e(0)}\\
Let $(\eta_\e)_{\e}$ be a non-negative mollifier, then for all 
 $f\in L\log L(\T^2)$ and $f\ge 0$, the function $f_{\e}=f\ast\eta_{\e}$ satisfies 
$$\displaystyle{\int_{\T^2} f_{\e}\ln f_{\e}  \rightarrow \int_{\T^2} f\ln f
\qquad\mbox{as}\qquad
\e\rightarrow 0.}$$ 

\end{lem}
\noindent For the proof see R. A. Adams 
\cite[Th 8.20]{Adams}.

\subsection{ {\it A priori} estimates }\label{EC:apriori}
In this subsection, we  show some $\varepsilon$-uniform estimates 
on the solutions of the system
(\ref{EC:eq:i:2})-(\ref{EC:initialapp}) obtained in Theorem
\ref{EC:theo:exip_regu}. These estimates will be used, on  the one hand
  to extend the solution in a global one and, on
the other hand in Subsection \ref{EC:preuv}, for ensuring by
compactness the passage to the limit as $\e$ tends to zero.

\noindent The first estimate concerns the physical entropy of the
system, and is a key result. It shows that in our model, the dislocation densities cannot be so
concentrated and then can 
be controlled.

\begin{lem}\label{EC:lemme:entro}{\bf(Entropy estimate)}\\ 
Let $\rho_0^{\pm}\in L^{2}_{loc}(\R^2)$.  If
$\rho^{\pm,\varepsilon}\in C^{\infty}(\mathbb R^2\times [0,T))$ are
solutions of the system (\ref{EC:eq:i:2})-(\ref{EC:initialapp}) and 
$\rho^{\pm,\varepsilon}(\cdot,t)$ satisfy $(H1)$, $(H2)$,
  $(H3)$ and $(H4)$, then

\begin{equation}\begin{array}{ll}\label{EC:entropyep}
\displaystyle{\int_{\T^2}}\sum_{\pm}\displaystyle{\frac{\partial{\rho}^{\pm,\e}}{\partial
      x_1}}\ln\left(\displaystyle{\frac{\partial{\rho}^{\pm,\e}}{\partial x_1}}\right)
+
\displaystyle{\int_0^t}\int_{\T^2}\left(R_1R_2\left(\displaystyle{
\displaystyle{\frac{\partial{\rho^\e}}{\partial x_1}}}\right)\right)^2\le 
\displaystyle{\int_{\T^2}\sum_{\pm}\displaystyle{\frac{\partial{\rho}^{\pm,\e}_0}{\partial
      x_1}}\ln\left(\displaystyle{\frac{\partial{\rho}_0^{\pm,\e}}{\partial x_1}}\right), }
\end{array}\end{equation}

\noindent where ${\rho}^{\e}={\rho}^{+,\e}-{\rho}^{-,\e}$.\\

\noindent In particular, there
exists a constant $C$  independent of $\e\in (0,1]$  such that
\begin{equation}\label{EC:keyest}
\left\|\displaystyle{\frac{\partial{\rho}^{\pm,\varepsilon}}{\partial
      x_1}}\right\|_{L^{\infty}\left((0,T); L\log L (\T^2)\right)}+
\left\|\displaystyle{\frac{\partial}{\partial
      x_1}}\left(R_1R_2\rho^{\e}\right)\right\|_{L^{2}\left(\T^2\times
    (0,T)\right )}\leq C
\end{equation}
\noindent with $C=C\left(\left\|\displaystyle{\frac{\partial{\rho_0}^{\pm}}{\partial x_1}}\right\|_{L\log
L(\T^2)}\right)$.
\end{lem}

\noindent {\bf Proof of Lemma \ref{EC:lemme:entro}:}

\noindent First of all, we denote 
$\theta^{\pm,\e}=\displaystyle{\frac{\partial{\rho}^{\pm,\varepsilon}}{\partial
    x_1}}$
and $N^{\pm}(t)=\displaystyle{\int_{\T^2}\theta^{\pm,\e}(t)\ln(\theta^{\pm,\e}(t))}.$

\noindent Using the fact that
$\rho^{\pm,\e}\in C^{\infty}(\mathbb R^2\times [0,T))$, we can derive $N(t)=N^+(t)+N^-(t)$
with respect to $t$, and since $\theta^{\pm,\e}>0$,
 we obtain:

$$
\displaystyle{\frac{d}{dt}N(t)}=\displaystyle{\int_{\T^2}\sum_{+,-}(\theta^{\pm,\e})_t\ln(\theta^{\pm,\e})+
\int_{\T^2}\sum_{+,-}(\theta^{\pm,\e})_t}.$$

\noindent Using system (\ref{EC:eq:i:2}) we see that the
second term is zero. Moreover, we get 

$$
\displaystyle{\frac{d}{dt}N(t)}=\displaystyle{\int_{\T^2}
\sum_{+,-}\left[\mp\left((R_1^2R_2^2\rho^{\e})\theta^{\pm,\e}\right)_{x_1}
+\varepsilon\Delta\theta^{\pm,\varepsilon}\right]\ln(\theta^{\pm,\e})}.$$

\noindent Integrating by part in $x_1$, we get
             
$$\begin{array}{ll}
\displaystyle{\frac{d}{dt}N(t)}
&=\displaystyle{\int_{\T^2}}\sum_{+,-}\left(\pm(R_1^2R_2^2\rho^{\e})\theta^{\pm,\e}\right)
\frac{\theta^{\pm,\e}_{x_1}}{\theta^{\pm,\e}}
-\varepsilon\sum_{+,-}\int_{\T^2}\frac{\left|\nabla\theta^{\pm,\e}\right|^2}{\theta^{\pm,\e}}\\
&=\displaystyle{\int_{\T^2}}\left(R_1^2R_2^2\rho^{\e}\right)\displaystyle{\frac{\partial{\theta^{\e}}}{\partial
    x_1}}
-\varepsilon\sum_{+,-}\int_{\T^2}\frac{\left|\nabla\theta^{\pm,\e}\right|^2}{\theta^{\pm,\e}}
\end{array}$$

\noindent where $\theta^{\e}=\theta^{+,\e}-\theta^{-,\e}$. 
We integrate also the first term  by part in $x_1$, and we deduce that 

$$\begin{array}{ll}
\displaystyle{\frac{d}{dt}N(t)}
&=-\displaystyle{\int_{\T^2}}\left(R_1^2R_2^2\theta^{\e}\right)\theta^{\e}
-\varepsilon\sum_{+,-}\int_{\T^2}\frac{\left|\nabla\theta^{\pm,\e}\right|^2}{\theta^{\pm,\e}}\\                 
             &=-\displaystyle{\int_{\T^2}\left(R_1R_2\theta^{\e}\right)^2
-\varepsilon\sum_{+,-}\int_{\T^2}\frac{\left|\nabla\theta^{\pm,\e}\right|^2}{\theta^{\pm,\e}}}
\leq0,
\end{array}$$

\noindent where we have used Lemma \ref{Riesz} (iii) and (iv) for the
second line.

\noindent Integrating in time,  we get

$$N(t)+\displaystyle{\int_{0}^t\int_{\T^2}\left(R_1R_2\theta^{\e}\right)^2}
\le N(0).$$

\noindent Which proves (\ref{EC:entropyep}). Moreover, we have

$$ N(0)\le
\displaystyle{\int_{\T^2}\sum_{+,-}\theta^{\pm,\e}(0)\log(e+\theta^{\pm,\e}(0))}.
$$

\noindent Since the initial data (\ref{EC:initial}) satisfies
$(H4)$, we deduce by  Lemma \ref{EC:e(0)} that there exists a positive constant
$C$ independent of $\e\in (0 1]$ such that
$$N(t)+\displaystyle{\int_{0}^t\int_{\T^2}\left(R_1R_2\theta^{\e}\right)^2}
\le C.$$

\noindent Let us now consider
$$N_1^{\pm}(t)
=\displaystyle{\int_{\T^2}
    \theta^{\pm,\e}(t)\log(e+\theta^{\pm,\e}(t))}.$$

\noindent We deduce, with another constant $C'>0$, that 

$$N_1^{+}(t)+N_1^{-}(t)+\displaystyle{\int_{0}^t\int_{\T^2}\left(R_1R_2\theta^{\e}\right)^2} 
\le C' $$

\noindent which joint to Lemma \ref{norm} implies (\ref{EC:keyest}). $\hfill\Box$   

\begin{rem}\label{EC:w12}{\bf($L^2$ estimate on the gradient of the vector field)}\\
We want to bound $\nabla\left(R_1^2R_2^2\rho^{\e}\right)$. To this end,
remark that  by the property of Riesz transform (see Lemma \ref{Riesz}
(iii)), we have

$$\displaystyle{\frac{\partial}{\partial
    x_1}}R_1^2R_2^2\rho^{\e}=R_1R_2\left(\displaystyle{\frac{\partial}{\partial
    x_1}}R_1R_2\rho^{\e}\right) \;\;\;\mbox{and}\;\;\;
\displaystyle{\frac{\partial}{\partial
    x_2}}R_1^2R_2^2\rho^{\e}=R_2^2\left(\displaystyle{\frac{\partial}{\partial
    x_1}}R_1R_2\rho^{\e}\right),$$ 

\noindent where those quantities involve $\displaystyle{\frac{\partial}{\partial
    x_1}}R_1R_2\rho^{\e}$ which is bounded in $L^{2}\left(\T^2\times
    (0,T)\right)$ by (\ref{EC:keyest}). Then using the fact the Riesz
transforms are continuous from $L^2$ onto itself  (see Lemma \ref{Riesz}
(i)), we deduce that 

\begin{equation}\label{grad}\left\|\nabla\left(R_1^2R_2^2\rho^{\e}\right)\right\|_{{L^{2}\left(\T^2\times
    (0,T)\right)}}\le C,\end{equation}

\noindent where  the constant $C$ is  independent on $\e$.
\end{rem}

\noindent We now present a  second a priori estimate.
\begin{lem}\label{EC:Ldeu}{\bf($L^2$ bound on the solutions)}\\
 Let $T>0$. Under the
 condition  $\rho_0^{\pm}\in L^{2}_{loc}(\R^2)$. If $\rho^{\pm,\e}\in
  C^{\infty}(\R^2\times [0,T))$
are solutions of system (\ref{EC:eq:i:2})-(\ref{EC:initialapp})  and
$\rho^{\pm,\e}(\cdot,t)$ satisfy $(H1)$,
  $(H2)$, $(H3)$ and $(H4)$, 
then there exists a constant $C_T$ independent of $\e\in(0 1]$, but depending on
$T$, such that:
  $$\left\|\rho^{\pm,\varepsilon,per}\right\|_{L^{\infty}((0,T);L^2(\T^2))}\leq C_T$$
 with $\rho^{\pm,\varepsilon,per}=\rho^{\pm,\varepsilon}-Lx_1$.
\end{lem}
\noindent {\bf Proof of Lemma \ref{EC:Ldeu}:}\\
\noindent Let $\T=\R/\Z$. We want to  bound 
$\displaystyle{m^{\pm,\e}(x_2,t)=\int_{\T}\rho^{\pm,\varepsilon,per}(x_1,x_2,t)dx_1}$. 
There is no problem of regularity since $\rho^{\pm,\e}\in
  C^{\infty}(\R^2\times [0,T))$. We integrate  equation
(\ref{EC:eq:i:2}) with respect to $x_1$, and we get

\begin{equation}\label{moyen}\begin{array}{ll}\displaystyle{\frac{\partial m^{\pm,\e}}{\partial
      t}}-\varepsilon\frac{\partial^2
 m^{\pm,\e}}{\partial
  x_2^2}
=
&\pm\displaystyle{\int_{ \T}(R_1^2R_2^2\displaystyle{\frac{\partial{\rho}^{\varepsilon}}{\partial
    x_1}})}(\rho^{\pm,\varepsilon,per}-
m^{\pm,\e}) dx_1
\pm m^{\pm,\e}\int_{\T}(R_1^2R_2^2\displaystyle{\frac{\partial{\rho}^{\varepsilon}}{\partial
    x_1}})dx_1\\
&\mp\displaystyle{ L_{\e}\int_{\T}(R_1^2R_2^2\rho^{\varepsilon})}dx_1,

\end{array}\end{equation}

\noindent where for the first line we have integrated by part, and
introduced the mean value $m^{\pm,\e}$. Therefore, using  that $\rho^{\e}$ is a $1$-periodic function in
$x_1$ and Lemma \ref{Riesz} (ii) and (iii), we deduce that 
$$\displaystyle{\int_{\T}(R_1^2R_2^2\rho^{\varepsilon})dx_1=0=\int_{\T}(R_1^2R_2^2
\displaystyle{\frac{\partial{\rho}^{\varepsilon}}{\partial
    x_1}})dx_1},$$

\noindent Equation (\ref{moyen}) is then equivalent to

\begin{equation}\label{EC:bornmoy}
\displaystyle{\frac{\partial m^{\pm,\e}}{\partial t}}-\varepsilon\frac{\partial^2 m^{\pm,\e}}{\partial
  x_2^2}
=\pm\displaystyle{\int_{\T}(R_1^2R_2^2\displaystyle{\frac{\partial{\rho}^{\varepsilon}}{\partial
    x_1}})}(\rho^{\pm,\varepsilon,per}-
m^{\pm,\e})dx_1={I^{\pm}(x_2,t)}.
\end{equation}
\noindent   We now show that
$I^{\pm}\in L^2(\T\times(0,T))$. Indeed, we have

$$\begin{array}{ll}\left\|I^{\pm}\right\|_{L^2(\T\times(0, T))}
&\le
\left\|\displaystyle{\int_{\T}(R_1^2R_2^2\displaystyle{\frac{\partial{\rho}^{\varepsilon}}{\partial
    x_1}})}(\rho^{\pm,\varepsilon,per}-
m^{\pm,\e})dx_1\right\|_{L^2(\T\times(0, T))}\\
\\
&\leq \left\|\rho^{\pm,\varepsilon,per}-
m^{\pm,\e}\right\|_{L^{\infty}(\T^2\times(0,T))}
  \left\|R_1^2R_2^2\displaystyle{\frac{\partial{\rho}^{\varepsilon}}{\partial
    x_1}}\right\|_{L^2(\T^2\times(0,T))}\\
\\
&\leq C
\end{array}$$
\noindent where for the last line we have used (\ref{grad}) and (Lemma
\ref{Riesz} (i)) to
bound  $\left\|R_1^2R_2^2\displaystyle{\frac{\partial{\rho}^{\varepsilon}}{\partial
    x_1}}\right\|_{L^2(\T^2\times(0,T))}$. Furthermore, the bound

 $$\left\|\rho^{\pm,\varepsilon,per}-
m^{\pm,\e}\right\|_{L^{\infty}(\T^2\times(0,T))}\le C$$ 

\noindent follows from (\ref{EC:moy}).\\

\noindent Multiplying (\ref{EC:bornmoy}) by $m^{\pm,\e}$ and
integrating in $x_2$, we get 
 
$$\displaystyle{\frac {1}{2}\frac {d}{dt}\|m^{\pm,\e}(\cdot, t)\|_{L^2(\T)}^2}+\e\left\|
\displaystyle{\frac{\partial }{\partial
    x_2}m^{\pm,\e}}(\cdot, t)\right\|_{L^2(\T)}^2=\int_{\T}(I^{\pm}m^{\pm,\e})(\dot,t).$$

\noindent Using Cauchy-Schwarz inequality on the right hand side, we deduce
that 

$$\displaystyle{\frac {1}{2}\frac {d}{dt}\|m^{\pm,\e}(\cdot, t)\|_{L^2(\T)}^2}
\le \|I^{\pm}(\cdot, t)\|_{L^2(\T)}.$$

\noindent We conclude to the result by integrating in  time.

 $\hfill\Box$

\begin{cor}\label{Field}{\bf($W^{1,2}$ estimate on the vector field)}\\
 Under the assumptions  $\rho_0^{\pm}\in
L^{2}_{loc}(\R^2)$. If $\rho^{\pm,\e}\in C^{\infty}(\R^2\times [0,T))$ are solutions of
the system (\ref{EC:eq:i:2})-(\ref{EC:initialapp}) and $\rho^{\pm,\varepsilon}(\cdot,t)$ satisfy $(H1)$, $(H2)$,
  $(H3)$ and $(H4)$, then there exists a
constant $C$ independent of  $\e$ such that:\\
$$\left\|R_1^2R_2^2\rho^{\e}\right\|_{{L^{2}((0,T);W^{1,2}(\T^2) )}}\le C,$$
\end{cor}

\noindent Using (\ref{grad}) and the fact that $R_1^2R_2^2\rho^{\e}$ is of
null average (see Lemma \ref{Riesz} (ii)) and applying
``Poincaré-Wirtinger inequality'', we can prove the result.

\noindent The following estimate will provide compactness in
time of the solution, uniform with respect to $\e$.

\begin{lem}{\bf (Duality estimate for the time derivative of the solution)}\\
\label{EC:lem:etem} Let $T>0$. Under the assumptions  $\rho_0^{\pm}\in
L^{2}_{loc}(\R^2)$. If $\rho^{\pm,\e}\in C^{\infty}(\R^2\times [0,T))$ are solutions of
the system (\ref{EC:eq:i:2})-(\ref{EC:initialapp}) and 
$\rho^{\pm,\varepsilon}(\cdot,t)$ satisfy $(H1)$, $(H2)$,
  $(H3)$ and $(H4)$, then 

\noindent i) For all  $\psi \in L^2((0,T);
W^{1,2}(\T^2))$, there exists a
constant $C$ independent of  $\e\in (0,1]$ such that:\\

$$\left|\int_{\T^2\times (0,T)}
\psi\displaystyle{R_1^2R_2^2}\left(\frac{\partial{\rho}^{\varepsilon}}{\partial
      t}\right)
\right|\leq
      C\|\psi\|_{L^{2}((0,T); W^{1,2}(\T^2))}$$
\noindent where
${\rho}^{\varepsilon}={\rho}^{+,\varepsilon}-{\rho}^{-,\varepsilon}$.
 
\noindent ii) For all  $\psi \in L^2((0,T);
W^{2,2}(\T^2))$, there exists a
constant $C_T$ independent of  $\e\in (0,1]$ such that:\\

$$\left|\int_{\T^2\times (0,T)}
\psi\displaystyle{\left(\frac{\partial{\rho}^{\pm,\varepsilon}}{\partial
      t}\right)}
\right|\leq
      C_T\|\psi\|_{L^{2}((0,T); W^{2,2}(\T^2))}.$$

\end{lem}
\noindent {\bf Proof of Lemma \ref{EC:lem:etem}:}

\noindent \underline{{\bf Proof of (i)}}: The idea is somehow to bound
$\displaystyle{R_1^2R_2^2\left(\frac{\partial{\rho}^{\varepsilon}}{\partial
t}\right)}$ using the available bounds on the right hand side of the equation
  (\ref{EC:eq:i:2}).
 
\noindent We will give a proof by duality.
 First of all, we   subtract the two  equations of system
 (\ref{EC:eq:i:2}) and we apply the Riesz transform  $R_1^2R_2^2$, to
 obtain that

\begin{equation}\label{EC:I}\displaystyle{R_1^2R_2^2\left(\frac{\partial{\rho}^{\varepsilon}}{\partial
      t}\right)}=
-\overbrace{\mathstrut
  R_1^2R_2^2\left((R_1^2R_2^2\rho^{\varepsilon})
\displaystyle{\frac{\partial {k^{\varepsilon}}}{\partial
x_1}}\right)}^{I_1}
+\overbrace{\mathstrut\varepsilon
R_1^2R_2^2\left(\Delta{\rho}^{\varepsilon}\right)}^{I_2}
\end{equation}

\noindent with $k^{\e}=\rho^{+,\e}+\rho^{-,\e}$. In what follows, we will prove that  for a function 
$\psi \in L^2((0,T); W^{1,2}(\T^2))$, we can bound
$J_i=\displaystyle{\int_{\T^2\times(0,T)}\psi I_i}$ for
$i=1, 2$.\\

\noindent \underline{{\bf Estimate of $J_2$}}: To estimate $J_2$,
we integrate by part, to get:
$$J_2=-\varepsilon\displaystyle{\int_{\T^2\times(0,T)}
\nabla{(R_1^2R_2^2\rho^{\varepsilon})}\cdot\nabla\psi}.$$

\noindent We deduce that for all $\e \in (0 1]$:

\begin{equation}\label{EC:I_2}\begin{array}{lll}

\left|J_2\right|
&\le \left\|R_1^2R_2^2\rho^{\varepsilon}\right\|_{L^{2}((0,T); W^{1,2}(\T^2))}
\|\psi\|_{L^2((0,T);W^{1,2}(\T^2))}\\
&\le C\|\psi\|_{L^2((0,T);W^{1,2}(\T^2))},
\end{array}\end{equation}

\noindent where we have used Corollary \ref{Field} in the last line.

\noindent \underline{{\bf Estimate of $J_1$}}: To control 
$J_1$, we rewrite it under the following form:
$$\displaystyle{\int_{\T^2\times(0,T)}
  \left[R_1^2R_2^2\left((R_1^2R_2^2\rho^{\varepsilon})\displaystyle{\frac{\partial 
{k^{\varepsilon}}}{\partial
x_1}}\right)\right]\psi}
=\displaystyle{\int_{\T^2\times(0,T)}
 \left((R_1^2R_2^2\rho^{\varepsilon})\displaystyle{\frac{\partial 
{k}^{\varepsilon}}{\partial
x_1}}\right) (R_1^2R_2^2\psi)}. $$
\noindent We use the fact that
\begin{enumerate}
\item[(i)]
$\left(R_1^2R_2^2\rho^\varepsilon\right)$ is bounded in 
  $L^{2}((0,T); W^{1,2}(\T^2))$ uniformly in $\e$ (by Corollary \ref{Field}),
\item[(ii)] $\displaystyle{\frac{\partial 
{k}^{\varepsilon}}{\partial x_1}}$
is bounded in $L^{\infty}((0,T);L \log L(\T^2))$, uniformly in $\e$ (by Lemma
\ref{EC:lemme:entro}).
\end{enumerate}

\noindent We deduce from this  and from Proposition 
\ref{EC:est_bil}, (with $f=R_1^2R_2^2\rho^\varepsilon$ and 
$g=\displaystyle{\frac{\partial {k}^{\varepsilon}}{\partial
    x_1}}$)  the following
estimate:

$$\begin{array}{lll}\left\|(R_1^2R_2^2\rho^\varepsilon )\displaystyle{\frac{\partial {k^\e}}{\partial
    x_1}}\right\|_{L^{2}((0,T); L\log^{\frac 12} L(\T^2))}
&\le 
C\|R_1^2R_2^2\rho^\varepsilon\|_{L^2((0,T);
  W^{1,2}(\T^2))}\left\|\displaystyle{\frac{\partial {k}^{\varepsilon}}
{\partial
    x_1}}\right\|_{L^{\infty}((0,T); L\log
  L(\T^2))}\\
\\
&\le C\left\|\displaystyle{\frac{\partial {k^{\varepsilon}}}{\partial
    x_1}}\right\|_{L^{\infty}((0,T); L\log
  L(\T^2))}\le C.
\end{array}$$

\noindent  We use Lemma \ref{EC:hold} (i), to deduce that
\begin{equation}\label{EC:I_1}\begin{array}{lll}|J_{1}|
&\le\left|\displaystyle{\int_{\T^2\times(0,T)}
 \left((R_1^2R_2^2\rho^{\varepsilon})\displaystyle{\frac{\partial 
{k}}{\partial
x_1}^{\varepsilon}}\right) (R_1^2R_2^2\psi)}\right|\\
\\
&\le
\left\|(R_1^2R_2^2\rho^\varepsilon) \displaystyle{\frac{\partial {k}}{\partial
    x_1}^{\varepsilon}}
\right\|_{L^{2}((0,T);L \log^{\frac 12}L(\T^2))}
\left\|R_1^2R_2^2\psi\right\|_{L^{2}((0,T); EXP_2(\T^2))}\\
\\
&\le
C\left\|R_1^2R_2^2\psi\right\|_{L^{2}((0,T);W^{1,2}(\T^2))} \le C\left\|\psi\right\|_{L^{2}((0,T);W^{1,2}(\T^2))}
\end{array}\end{equation}
\noindent where we have used the Trudinger inequality (see Lemma
\ref{EC:trud}) in the third line and the fact that Riesz
transforms are continuous from $L^{2}$ onto itself in the last
line (see Lemma \ref{Riesz} (i)).

\noindent Finally,  collecting (\ref{EC:I_1}) and (\ref{EC:I_2}) 
 together with (\ref{EC:I}) and the definitions of $J_i$, for $i=1,2$, we get that
there exists a constant $C$ independent of $\e$ such that

$$\left|\displaystyle{\int_{\T^2\times(0,T)}}\psi R_1^2R_2^2
(\displaystyle{\frac{\partial{\rho}^{\varepsilon}}{\partial
    t}})\right|
\leq C\|\psi\|_{L^2((0,T);W^{1,2}(\T^2))}.
$$

\noindent \underline{{\bf Proof of ii)}}: The proof of (ii) is similar to that of (i). 
The only difference is that we integrate by part the viscosity term
 twice and use the estimate of Lemma \ref{EC:Ldeu}.
$\hfill\Box$

\begin{rem}\label{EC:w(-21)}{\bf($W^{-1,2}$ and $W^{-2,2}$ estimate)}\\
Let $W^{-1,2}(\T^2)$ be the dual space of $W^{1,2}(\T^2)$. By  
  point (i)  of the previous lemma, we deduce that
  there exists a constant $C$ independent of  $\e$, such that

$$\left\|\displaystyle{R_1^2R_2^2}\left(\frac{\partial{\rho}^{\varepsilon}}{\partial
      t}\right)\right\|_{L^{2}\left((0,T);W^{-1,2}(\T^2)\right)}\leq C.$$

\noindent However,  the point (ii) controls the time derivative of the solution in
$L^{2}\left((0,T);W^{-2,2}(\T^2)\right)$, where $W^{-2,2}(\T^2)$
is the dual space of $W^{2,2}(\T^2)$. This control will allows us later 
to recover the initial conditions in the limit as $\e$ goes to zero.
\end{rem}

\begin{theo}\label{EC:theo:exip1}{\bf (Global existence)}\\
For all $T>0$, $\e\in (0,1]$ and for all initial data $\rho_0^{\pm}\in
L^{2}_{loc}(\R^2)$ satisfying $(H1)$, $(H2)$, $(H3)$ and $(H4)$, the system
(\ref{EC:eq:i:2})-(\ref{EC:initialapp}) admits a solution
$\rho^{\pm,\e}\in  C^{\infty}(\R^2\times [0,T))$. Moreover,
$\rho^{\pm,\e}(\cdot,t)$  satisfies  $(H1)$, $(H2)$
and $(H3)$ for all  $t\in (0,T)$ and the estimates
given in Lemmata \ref{EC:lemme:entro}, \ref{EC:Ldeu}, \ref{EC:lem:etem}
and  Corollary \ref{Field}.   
\end{theo}

\noindent   Before going  into the proof, we need the following lemma.

\begin{lem}\label{EC:estsemi2}{\bf($W^{1,\frac 32}$ estimate)}\\
For all initial data $\rho_0^{\pm}\in L^{2}_{loc}(\R^2)$
satisfying  $(H1)$ and $(H2)$,  if 
$\rho^{\pm,\e,per}\in C^{\infty}(\T^2\times [0,T))$, are solutions of
the Mild integral problem (\ref{EC:eq:i:6}), then there exists a constant 
$C=C(\e,L)$ such that,
{\small\begin{displaymath}\|\rho^{\pm,\e,per}\|_{L^{\infty}((0,T);W^{1,\frac 32}(\T^2))}\le
B_0^{\pm}+
CT^{\frac
  {1}{24}}\|R_1^2R_2^2\rho^{\e}\|_{L^{\infty}((0,T);L^{8}(\T^2))}
\left(\left\|\displaystyle{\frac{\partial{\rho}^{\pm,\varepsilon}}{\partial
    x_1}}\right\|_{L^{\infty}((0,T);L^1(\T^2))}+1\right),\end{displaymath}}
where $B_0^{\pm}=\|\rho_0^{\pm,\e,per}\|_{W^{1,\frac 32}(\T^2)}$.
\end{lem}

\noindent {\bf Proof of Lemma \ref{EC:estsemi2}:}\\
\noindent If we denote 
$\rho_v^{\e}=(\rho^{+,\varepsilon,per},\rho^{-,\varepsilon,per})$ and
$\rho_{0,v}^{\e}=(\rho^{+,\e,per}_0,\rho^{-,\e,per}_0)$, then we have
shown that $\rho_v^{\e}$ satisfies (\ref{EC:eq:point}), using (\ref{EC:est33})
with $u=v=\rho_v^{\varepsilon}$ , we get,

 $$\begin{array}{ll}\|B(\rho_v^{\varepsilon},\rho_v^{\varepsilon})(t)\|_{(W^{1,\frac 32}(\T^2))^2}
&\leq\displaystyle{\int_0^t}\left\|
S_{\e}(t-s)\left(\left(R^2_1R^2_2\rho^{\varepsilon}(s)\right)\displaystyle{\frac{\partial{\rho_v^{\varepsilon}}}{\partial

    x_1}(s)}\right)ds\right\|_{(L^{4}(\T^2))^2}\\
\\
&+
\displaystyle{\int_0^t}\left\|\nabla
S_{\e}(t-s)\left(\left(R^2_1R^2_2\rho^{\varepsilon}(s)\right)\displaystyle{\frac{\partial{\rho_v^{\varepsilon}}}{\partial
    x_1}(s)}\right)ds\right\|_{(L^{\frac 32}(\T^2))^2}.
\end{array}$$

\noindent We use now Lemma \ref{EC:estsemi} (i) with
$r=4,q=\frac {24}{5},p=1$ to estimate the first term, and Lemma \ref{EC:estsemi}
(ii) with  $r=\frac 32,q=8,p=1$ to  estimate the second  term. It gives
for $t\in (0, T)$, that,\\

{\small\begin{displaymath}\begin{array}{ll}
\hspace{-1em}\|B(\rho_v^{\varepsilon},
\rho_v^{\varepsilon})(t)\|_{(W^{1,\frac 32}(\T^2))^2}
&\hspace{-0.7em}\leq
C\displaystyle{\int_0^t\frac{1}{(t-s)^{\frac{23}{24}}}}
\left\|R^2_1R^2_2\rho^{\varepsilon}(s)\right\|_{L^8(\T^2)}
\left\|\displaystyle{\frac{\partial{\rho_v^{\varepsilon}}}{\partial
      x_1}(s)}\right\|_{(L^1(\T^2))^2
}ds\\
\\
&\hspace{-1em}\leq C\displaystyle{\sup_{0\leq
    s<T}}\left(\left\|R^2_1R^2_2\rho^{\varepsilon}(s)\right\|_{L^8(\T^2)}\right)
\sup_{0\leq
  s<T}\left(\left\|\displaystyle{\frac{\partial{\rho_v^{\varepsilon}}}{\partial
      x_1}(s)}\right\|_{(L^1(\T^2))^2}\right)
\displaystyle{\int_0^t\frac{1}{(t-s)^{\frac{23}{24}}}}\cdot
\end{array}\end{displaymath}}
\noindent That leads,
 \begin{equation}\label{EC:eq:bi1}
\|B(\rho_v^{\varepsilon},\rho_v^{\varepsilon})
\|_{L^{\infty}((0,T);(W^{1,\frac 32}(\T^2))^2)}\leq
C T^{\frac {1}{24}}
\|R^2_1R^2_2\rho^{\varepsilon}\|_{L^{\infty}((0,T);L^8(\T^2))}
\left\|\displaystyle{\frac{\partial{\rho_v^{\varepsilon}}}{\partial
      x_1}}\right\|_{L^{\infty}((0,T); (L^1(\T^2))^2)}\cdot
\end{equation}
\noindent Similarly, we show that,
\begin{equation}\label{EC:eq:li1}\|A(\rho_v^{\varepsilon})\|_{L^{\infty}((0,T);W^{1,\frac 32}(\T^2))}\leq
C T^{\frac {1}{24}} \|R^2_1R^2_2\rho^{\varepsilon}\|_{L^{\infty}((0,T);L^8(\T^2))}.
\end{equation}

\noindent By using  (\ref{EC:eq:bi1}), (\ref{EC:eq:li1}) and 
(\ref{EC:eq:do}), and the equation ((\ref{EC:eq:point})) we get the proof. 

  $\hfill\Box$

\noindent{\bf Proof of Theorem \ref{EC:theo:exip1}:}\\
We argue by contradiction. 
Suppose that there exists a maximum time $T_{max}$ such
that we have the existence of solutions of 
(\ref{EC:eq:i:2})-(\ref{EC:initialapp}) in $
C^{\infty}(\R^2\times[0,T_{max}))$.\\    

\noindent For $\delta>0$, we reconsider the system  (\ref{EC:eq:i:2}) with
the initial data 
$$\rho^{\pm,\e}_{\delta,max}=\rho^{\pm,\e}(x,T_{max}-\delta).$$

\noindent We reapply for the second time, the proof of Theorem
\ref{EC:theo:exip_regu}, we deduce that there exists a time  
$$T^{\star}_{\delta,max}(\|\rho^{\pm,\e,per}_{\delta,max}\|_{W^{1,\frac
    32}(\T^2)},L,\e)>0,
\;\quad\mbox{where}\;\quad\rho^{\pm,\e,per}_{\delta,max}=\rho^{\pm,\e}_{\delta,max}-Lx_1,$$ 
such that the system  (\ref{EC:eq:i:2})-(\ref{EC:initialapp}) admits solutions
defined until,
$$T_0=(T_{max}-\delta)+T^{\star}_{\delta,max}.$$
 
\noindent Moreover,  from Lemmata \ref{EC:lem:ebmo}
  \ref{Riesz}  (v) and \ref{Riesz} (i) with  $p=8$, we can deduce  easily that $R_1^2R_2^2(\rho^\e)$  is bounded on 
 $L^{\infty}((0,T),L^8(\T^2))$. Now,  by Lemmata  \ref{EC:estsemi2} and \ref{EC:lemme:entro}, 
 we know that $\rho^{\pm,\e,per}_{\delta,max}$ are $\delta$-uniformly bounded in $W^{1,\frac 32}(\T^2)$. By using (\ref{EC:tstar}), 
 we deduce that there exists a constant
$C(\e,T_{max},L)>0$ independent of $\delta$ such that 
$T^{\star}_{\delta,max}\ge C>0$. Then $
\displaystyle{\liminf_{\delta\rightarrow 0}T^{\star}_{\delta,max}\ge C>0}$.
 Hence $T_0>T_{max}$ which gives the contradiction. 
 
 $\hfill\Box$

\section{Existence of solutions for the system 
(\ref{EC:eq:i:1})-(\ref{EC:initial})}

In this section, we will prove that the system
(\ref{EC:eq:i:1})-(\ref{EC:initial}) admits solutions $\rho^{\pm}$ in the
distributional sense.  They are the limits  when  $\e\rightarrow
0$ of the solution $\rho^{\pm,\e}$ given in
Theorem \ref{EC:theo:exip1}. To do this, we
will justify the passage  to the limit as $\e$ tends to $0$ in the system
(\ref{EC:eq:i:2})-(\ref{EC:initialapp}), using some 
compactness arguments.\\    

\subsection{Preliminary results}
Before proving the main theorem, let us recall some well known results.

\begin{lem}\label{EC:lem:comp}{\bf (Trudinger compact embedding)}\\
\noindent The following injection (see N. S. Trudinger \cite{Trud}):
$$W^{1,2}(\T^2)\hookrightarrow EXP_\beta(\T^2),$$
\noindent is compact, for all $1\le \beta <2$.
\end{lem}
\noindent For the proof of this lemma see also R. A. Adams \cite[Th
8.32]{Adams}.

\begin{lem}\label{EC:simo}{\bf (Simon's Lemma)}\\
 Let $X$, $B$, $Y$ three  Banach spaces, where $X\hookrightarrow B$ with
 compact embedding and $B\hookrightarrow Y$ with continuous
 embedding. If $(\rho^n)_n$ is a sequence such that

$$\|\rho^n\|_{L^q((0,T); B)}+\|\rho^n\|_{L^1((0,T); X)}+
\left\|\displaystyle{\frac{\partial \rho^n}{\partial
    t}}\right\|_{L^1((0,T); Y)}\le C,
$$
\noindent  where $q>1$ and $C$ is a constant independent of $n$, 
then $(\rho^n)_n$ is relatively compact in $L^p((0,T); B)$ for all $1\le
p<q$.

\end{lem}
\noindent For the proof, see J. Simon \cite[Th 6, Page 86]{SI87}.

\noindent In order to show the global existence of system (\ref{EC:eq:i:1}) in
Subsection \ref{EC:preuv}, we will apply this lemma in the particular cases
where $B=EXP_\beta(\T^2)$, $X=W^{1,2}(\T^2)$ and $Y=W^{-1,2}(\T^2)$, for
$1<\beta<2$.

\begin{lem}\label{EC:weak}{\bf (Weak star topology in $L\log L$)}\\
Let $E_{exp}(\T^2)$ be the closure in $EXP(\T^2)$ of the
space of functions  bounded on $\T^2$. Then $E_{exp}(\T^2)$  is a separable
Banach  space which verifies, \\

\noindent i) \centerline{$\mbox{$L\log
L(\T^2)$ is the dual space of $E_{exp}(\T^2)$.}$} \\

\noindent ii) \centerline{$\mbox{$EXP_\beta(\T^2)
\hookrightarrow E_{exp}(\T^2)\hookrightarrow EXP(\T^2)$ for all $\beta > 1$.}$}
\end{lem}
\noindent For the proof, see R. A. Adams \cite[Th 8.16, 8.18, 8.20]{Adams}.

\subsection {Proof of Theorem \ref{EC:theo:exi}}\label{EC:preuv}

\noindent  Let us fix any $T>0$. For any $\e\in (0,1]$, we are
considering the solution $\rho^{\pm,\e}$ of
(\ref{EC:eq:i:2})-(\ref{EC:initialapp}) given in Theorem
\ref{EC:theo:exip1} on $\R^2\times (0,T)$. First,
by Lemma \ref{EC:Ldeu} we know that, the
periodic part of the solutions, denoted by $\rho^{\pm,\e,per}$ 
are $\e$-uniformly bounded in
$L^{2}(\T^2\times(0,T))$. Hence,  as $\e$
goes to zero, we  can extract a subsequence   still denoted
 by $\rho^{\pm,\e,per}$, that converges weakly  in
$L^{2}(\T^2\times(0,T))$ to some limit  $\rho^{\pm,per}$. 
Then we want to  prove that $\rho^{\pm}=\rho^{\pm,per}+Lx_1$ are solutions of the
system (\ref{EC:eq:i:1})-(\ref{EC:initial}). Indeed, since the passage to the
limit in the linear term is trivial in $\D'(\T^2\times(0,T))$, it
suffices to pass to the limit in the non-linear term

\begin{equation}\label{bil}(R_1^2R_2^2\rho^\varepsilon)\displaystyle{\frac{\partial
      \rho^{\pm,\varepsilon}}{\partial
    x_1}}.\end{equation}

\noindent \underline{{\bf Step 1 (compactness of
    $(R_1^2R_2^2\rho^\varepsilon)$)}}: Now notice that:

\noindent $\bullet$ From Corollary \ref{Field} we know that the term
  $\left(R_1^2R_2^2\rho^\varepsilon\right)$ is $\e$-uniformly bounded in 
  $L^{2}((0,T); W^{1,2}(\T^2))$. Then  it is in
particular $\e$-uniformly  bounded in $L^{1}((0,T); W^{1,2}(\T^2))$.

\noindent $\bullet$ From the  previous point and Lemma \ref{EC:lem:comp}, we know that 
$\left(R_1^2R_2^2\rho^\varepsilon\right)$ is also
$\e$-uniformly bounded in $L^{2}((0,T); EXP_\beta(\T^2))$ for all $1\le
\beta <2$.

\noindent $\bullet$ From Lemma  \ref{EC:lem:etem}, the term 
$R^2_1R^2_2(\displaystyle{\frac{\partial{\rho}^{\varepsilon}}{\partial
    t}})$ is $\e$-uniformly bounded in $L^{2}((0,T);W^{-1,2}(\T^2))$ and
then in $L^{1}((0,T);W^{-1,2}(\T^2))$.\\

\noindent Collecting this, we get that there exists a
constant $C$  independent on $\e$ such that $\bar{\rho}^{\e}=R_1^2R_2^2\rho^\varepsilon$
satisfies for some $1<\beta<2$
$$\left\|\bar{\rho}^{\e}\right\|_{L^{2}((0,T); EXP_{\beta}(\T^2))}+
\left\|\bar{\rho}^{\e}\right\|_ {L^{1}((0,T); W^{1,2}(\T^2))}+
\left\|\displaystyle{\frac{\partial{\bar{\rho}}^{\e}}{\partial
      t}}\right\|_{L^1((0,T); W^{-1,2}(\T^2))}
\leq C.$$

\noindent Then Lemma \ref{EC:simo} joint to Lemma \ref{EC:lem:comp}, with  $B=EXP_\beta(\T^2)$, $X=W^{1,2}(\T^2)$
and $Y=W^{-1,2}(\T^2)$, shows the relative compactness of 
$\left(R_1^2R_2^2\rho^\varepsilon\right)$
in $L^{1}((0,T); EXP_{\beta}(\T^2))$, and then using Lemma \ref{EC:weak}, 	
we deduce the compactness in $L^{1}((0,T); E_{exp}(\T^2))$.\\

\noindent \underline{{\bf Step 2 (weak-$\star$ convergence of 
$\displaystyle{\frac{\partial{\rho}^{\pm,\varepsilon}}{\partial
      x_1}}$)}}:
\noindent By Lemma \ref{EC:lemme:entro}, we have that
$\displaystyle{\frac{\partial{\rho}^{\pm,\varepsilon}}{\partial
      x_1}}$ is $\e$-uniformly bounded in
  $L^{\infty}((0,T); L\log L(\T^2))$ which is the dual of $L^{1}((0,T);
  E_{exp}(\T^2))$ by Lemma \ref{EC:weak}. 
Then, this term converges weakly-$\star$ in $L^{\infty}((0,T); L\log L(\T^2))$ toward
$\displaystyle{\frac{\partial{\rho}^{\pm}}{\partial
      x_1}}$.  
That enables us to pass to the limit 
in the bilinear term (\ref{bil}) in the sense 

$$L^{1}((0,T); E_{exp}(\T^2))-strong\;
\times\; L^{\infty}((0,T); L\log L(\T^2))-weak-\star.$$

\noindent which shows that
 
$$(R_1^2R_2^2\rho^\varepsilon)\displaystyle{\frac{\partial \rho^{\pm,\varepsilon}}{\partial
    x_1}}\rightarrow
(R_1^2R_2^2\rho)\displaystyle{\frac{\partial \rho^{\pm}}{\partial
    x_1}}\;\;\; \mbox{in $\D'(\T^2\times(0,T))$}.$$

\noindent In what precedes, we have shown that $\rho^{\pm}$
are solutions of the system (\ref{EC:eq:i:1}).\\

\noindent \underline{{\bf Step 3 (conclusion)}}:
\noindent Passing to the limit in the estimates of Lammata \ref{EC:lemme:entro},
\ref{EC:Ldeu}, \ref{EC:lem:etem} and Corollary \ref{Field}, we get in particular by Lemma
\ref{EC:e(0)}, the entropy
estimates (\ref{EC:entropy}) and $(E1)$, $(E2)$, $(E4)$, $(E5)$. At this stage
we remark that, by Proposition \ref{EC:est_bil} that 
$$\displaystyle{\frac{\partial \rho^{\pm}}{\partial
    t}}=(R_1^2R_2^2\rho)\displaystyle{\frac{\partial \rho^{\pm}}{\partial
    x_1}} \in  L^2((0,T);L\log^{\frac 12}L(\T^2)),$$

\noindent  and then
  ${\rho}^{\pm,per}\in C([0,T);L\log^{\frac 12}L(\T^2))$, which proves
  $(E3)$.\\

\noindent Since the function ${\rho}^{\pm,per,\e}(\cdot,t)$ satisfy
$(H1)$, $(H2)$, $(H3)$, $(H4)$ (see Theorem \ref{EC:theo:exip1}) by
passing in the limit $\e \rightarrow 0$, we can see that the limit 
function ${\rho}^{\pm,per}(\cdot,t)$ reserves the same
assumptions $(H1)$, $(H2)$, $(H3)$, $(H4)$. \\

\noindent It remains to prove that $\rho^{\pm}$ satisfies  the initial
conditions (\ref{EC:initial}). Indeed, from the estimates on
$\rho^{\pm,\e,per}$ given by Lemma \ref{EC:Ldeu} and
$\displaystyle{\frac{\partial{\rho}^{\pm,\e}}{\partial t}}$ given by
Lemma  \ref{EC:lem:etem} (ii), we can prove easily, that 

$$\|\rho^{\pm,\e,per}(t)-\rho^{\pm,\e,per}_0\|_{W^{-2,2}(\T^2)}\le
C_Tt^{\frac 12}.$$
where $C_T$ is constant independent of $\e$. Hence we can pass to the
limit $\e \rightarrow 0$, which this implies  in particular that 
$\rho^{\pm,per}(\cdot,0)=\rho_0^{\pm,per}$ in
$\D'(\R^2)$. $\hfill\Box$

\begin{rem}{$\;$}\\ 
In our proof, we have indirectly used a kind of compensated
  compactness technic for Hardy spaces. Nevertheless in our case, we do
  not have enough regularity  to do so.
\end{rem}

\section{Acknowledgements }
The second author would like to thank Y. Meyer, F. Murat and L. Tartar
 for fruitful remarks that helped in the preparation of the paper, and  
 H. Ibrahim for carefuly reading it. The authors also would like to
 thank the referee who helped to improve drastically  the presentation of the paper.
This work was partially
supported by the contract JC 1025 ``ACI,
jeunes chercheuses et jeunes chercheurs'' (2003-2007), the program ``PPF, programme pluri-formations mathématiques
financières et EDP'', (2006-2010), Marne-la-Vall\'ee University and École Nationale
des Ponts et Chaussées, and by the project ANR MICA (``Mouvements d'interfaces,
calcul et applications'').

 \bibliographystyle{siam}
 \bibliography{biblio}

\end{document}